\documentstyle[aps,preprint,epsf]{revtex}

\begin{document}  
\draft

\title{Non-monotonic variation with salt concentration of the 
second virial coefficient in protein solutions} 
 
\author {E. Allahyarov$^1$, H. L\"{o}wen$^2$, J.P. Hansen$^3$, and A.A. Louis$^3$} 
 
\address{$^1$Institut f\"ur Festk\"orperforschung, Forschungszentrum
  J\"ulich, D-52425 J\"ulich, Germany}
 \address{$^2$Institut f\"{u}r Theoretische Physik II, 
Heinrich-Heine-Universit\"{a}t D\"{u}sseldorf, D-40225 D\"{u}sseldorf, 
Germany} 
\address{$^3$Department of Chemistry, Lensfield Rd, Cambridge CB2 1EW, 
UK} 
\maketitle 
\begin{abstract} 
The osmotic virial coefficient $B_2$ of globular protein solutions is 
calculated as a function of added salt concentration at fixed pH by 
computer simulations of the ``primitive model''.  The salt and 
counter-ions as well as a discrete charge pattern on the protein   
surface are explicitly incorporated.  For parameters roughly 
corresponding to lysozyme, we find that $B_2$ first decreases with 
added salt concentration up to a threshold concentration, then 
increases to a maximum, and then decreases again upon further raising 
the ionic strength.  Our studies demonstrate that the existence of a 
discrete charge pattern on the protein surface profoundly influences 
the effective interactions and that non-linear Poisson Boltzmann and
Derjaguin-Landau-Verwey-Overbeek (DLVO) theory fail for large ionic strength.  The observed 
non-monotonicity of $B_2$ is compared to experiments. Implications for 
protein crystallization are discussed. 
\end{abstract} 

\section{introduction}

Interactions between proteins in aqueous solutions determine their 
collective behavior, in particular their aggregation, their 
complexation with other macromolecules, and ultimately their phase 
behavior, including phase separation, precipitation and 
crystallization.  Any theoretical analysis of the properties of 
protein solutions must rely on a reliable understanding of their 
interactions.  A good example is provided by the control of protein 
crystallization, which is an essential prerequisite for the 
determination of protein structure by X-ray 
diffraction\cite{Durbin,Tardieu}.  While at present protein 
crystallization is still mostly achieved experimentally by ``trial and 
error'', and on the basis of a number of empirical rules\cite{Geor94}, 
there is clearly a need for a more fundamental understanding of the 
mechanisms controlling protein crystallization, and this obviously 
requires a good knowledge of the forces between protein molecules in 
solution, and of their dependence on solution conditions, including pH 
and salt concentration\cite{Durbin,protein,protein4,Dill99}. 
 
Protein interactions have various origins, and one may conveniently 
distinguish between direct and induced (or effective) contributions. 
Direct interactions include short-range repulsive forces, which 
control steric excluded volume effects, reflecting the shape of the 
protein, van der Waals dispersion forces, and electrostatic forces 
associated with pH-dependent electric charges and higher electrostatic 
multipoles carried by the protein residues\cite{Tamashima}.  Other, 
effective, interactions depend on the degree of coarse-graining in the 
statistical description and result from the tracing out of microscopic 
degrees of freedom associated with the solvent and added electrolyte, 
i.e. the water molecules and microions.  Tracing out the solvent 
results in hydrophobic attraction and hydration forces, while 
integrating over microion degrees of freedom leads to screened 
electrostatic interactions between residues, the range of which is 
controlled by the Debye screening length, and hence by electrolyte 
concentration. 
 
However, while coarse-graining through elimination of microscopic 
degrees of freedom, leading to state-dependent effective interactions 
is a priori a reasonable procedure to describe highly asymmetric 
colloidal systems, where particles have diameters of typically 
hundreds of nm and carry thousands of elementary charges, this is 
obviously less justified for the much smaller and less charged 
proteins.  In particular the assumption of uniformly charged colloid 
surfaces, leading to spherically symmetric, screened interactions 
between the electric double layers around colloid particles, as 
epitomized by the classic DLVO (Derjaguin-Landau-Verwey-Overbeek) 
potential\cite{Verwey}, ceases to be a reasonable approximation at the 
level of nanometric proteins carrying typically of order 10 elementary 
charges.  The reason is that length scales which are widely separated 
in colloidal assemblies, become comparable in protein solutions, while 
the discreteness of charge distributions on proteins can no longer be 
ignored, since the distance between two charged residues on the 
protein surface is no longer negligible compared to the protein 
diameter.  Thus, electrostatic, as well as other (e.g. hydrophobic) 
interactions are much more specific in proteins, and must be 
associated with several interaction sites, rather than merely with the 
centers of mass as is the case for (spherical) colloidal particles. 
 
Another very important distinction between colloids and protein 
solutions is that the forces between the former may be measured 
directly, using optical means\cite{Kepl94,Lars97,Verm98}, while 
interactions between proteins can only be inferred indirectly, from 
measurements by static light scattering of the osmotic equation of 
state which, at sufficiently low concentrations, yields the second 
osmotic virial coefficient $B_2$\cite{Geor94,Rose96,Moon,Moon2}, the 
main focus of the present paper.  The variation of $B_2$ with solution 
conditions yields valuable information on the underlying effective 
pair interactions between proteins.  Moreover it was shown empirically 
by George and Wilson\cite{Geor94} that there is a strong correlation 
between the measured values of $B_2$ and the range of solution 
conditions which favor protein 
crystallization\cite{Rose96,Rose95,Vlie00}.  Only if the measured 
value of $B_2$ falls within a well defined ``slot'' can 
crystallization be achieved.  If $B_2$ is too large, repulsive 
interactions predominate, leading to slow crystallization rates.  On 
the other hand if $B_2$ is highly negative, strong attractions lead to 
amorphous aggregation. 
 
The correlation between $B_2$ and crystallization may be rationalized 
by noting that protein crystals generally coexist with a fairly dilute 
protein solution, the thermodynamic properties (and in particular the 
free energy) of which are essentially determined by $B_2$. 
Coexistence between a dense solid phase and a dilute fluid phase is 
generally a signature of a very short-ranged attraction between 
particles as compared to their 
diameter\cite{Vlie00,Hage94,Muschol,Malfois}. 
 
For such short-ranged attractive interactions, the phase-separation 
into dilute and concentrated proteins solutions expected on the basis 
of a mean-field, van der Waals theory, is in fact pre-empted by the 
freezing transition, i.e. the critical (or ``cloud'') point lies below 
the freezing line.  The critical fluctuations associated with this 
metastable cloud point may lead to a significant enhancement of the 
crystal nucleation rate\cite{tenWolde}, while the position of the 
cloud point in the concentration-temperature plane is strongly 
correlated with the virial coefficient $B_2$\cite{Vlie00}. 
 
The present paper focuses on the variation of $B_2$ with ionic 
strength of added salt.  This is a particularly important issue since 
``salting out `` of protein solutions is one of the standard methods 
used to induce crystallization.  An increase in salt concentration 
reduces the screening length and hence the electrostatic repulsion, 
allowing  short range attractive forces (e.g. of hydrophobic or 
van der Waals origin) to come into play which will ultimately trigger 
nucleation.  Recent experiments point to a non-monotonic variation of 
$B_2$ with increasing ionic strength\cite{Tessier,SunWalz}, or to a pronounced 
shoulder in the $B_2$ versus ionic strength curve\cite{Rosenbaum} in 
lysozyme solutions.  Closely related findings are the observation of a 
non-monotonic cloud point\cite{Wu,Broide,Taratuta}, and of a minimum 
in the solubility of lysozyme with increasing salt 
concentration\cite{Arakawa}; the solubility is obviously related to 
the osmotic virial coefficient\cite{Haas}.  Similarly, the attractive 
interaction parameter $\lambda$, which controls the variation of the 
measured protein diffusion coefficient $D$ with volume fraction, was 
found to exhibit a sharp minimum upon an increase of ionic strength of 
lysozyme solutions\cite{Grigsby}; again, this interaction parameter 
strongly correlates with $B_2$\cite{Guo,Bonnete}. 
 
Traditional models for the protein--protein interaction cannot easily 
reproduce such non-monotonic behavior of $B_2$ or related quantities. 
The ``colloidal'' approach based on spherical particles interacting 
via the screened Coulomb DLVO potential\cite{Verwey} can only predict 
a monotonic decrease of $B_2$ with ionic 
strength\cite{protein4,Piazza3}.  The same is true of 
models\cite{protein4,Rose96,Rose95} accounting for short-range 
attractions via Baxter's ``adhesive sphere'' 
representation\cite{baxter}.  In these models, which assume central 
pair-wise interactions, $B_2$ reduces to a simple integral of the 
Mayer function associated with the spherically symmetric 
potential\cite{Poon,protein3}.  More recent calculations account for 
the asymmetric shape of proteins\cite{SunWalz,Neal}, or include 
several ``sticky'' sites at the surface of the 
protein\cite{Sear,Benedek}. 
 
In these traditional calculations, electrostatic interactions between 
proteins and microions are routinely treated within mean-field 
Poisson-Boltzmann (PB) theory, generally in its linearized version (as 
is the case for the classic DLVO potential).  However, as explained 
earlier, all relevant length scales (i.e. protein diameter, mean 
distance between charged sites on the protein surface, and between co 
and counterions, as well as the Debye screening length) are comparable 
in protein solutions, so that the discrete nature of both the 
interaction sites, and of the co and counterions, can no longer be 
ignored.  Moreover, Coulomb correlations are expected to be enhanced 
on protein length scales and may lead to strong deviations from the 
predictions of PB theory, which have recently been shown to induce 
short-range attractions, even between much larger colloidal 
particles\cite{Wu,AllahyarovPRL,Linse,Pincus,HansenLoewen}. 
 
The present paper takes into account the discrete nature of the 
microions within a ``primitive model'' description of the electrolyte, 
and presents results of Molecular Dynamics calculations of the 
equilibrium distribution of co and counterions around two proteins and 
of the resulting osmotic virial coefficient $B_2$.  Two models of the 
charge distribution on the surface of the spherical proteins will be 
considered.  In the colloid-like model the charge is assumed to be 
uniformly distributed over the surface, while in the discrete charge 
model, the charges are attached to a small number of interaction 
sites.  The latter model will be shown to lead to a distinctly 
non-monotonic variation of $B_2$ with ionic strength, as observed 
experimentally. 
 
The paper is organized as follows: The model and key physical 
quantities are introduced in section II.  Simulation details are 
described in section III.  Results of the simulations are presented and 
discussed in section IV, while conclusions are summarized in section 
V.  A preliminary account of parts of the results was briefly reported 
elsewhere\cite{ourEPL}.

\section{Models, effective forces and second virial coefficient} 
 
The globular proteins under consideration are modeled as hard spheres 
of diameter $\sigma_p$, carrying a total (negative) charge 
$-Ze$. Within a ``primitive model'' representation\cite{Frie62}, the 
molecular granularity of the aqueous solvent is ignored, and replaced 
by a continuum of dielectric permittivity $\epsilon$, while the 
monovalent counterions and salt ions are assumed to have equal 
diameters $\sigma_s$ and charges $\pm e$. 
 
Two models are considered for the charge distribution on the surface 
of the protein.  In the ``smeared charge model'' (SCM), the total 
charge $-Ze$ is assumed to be uniformly distributed over the spherical 
surface, which is the standard model for charge-stabilized colloidal 
suspensions\cite{Wu,AllahyarovPRL,Linse,Pincus,HansenLoewen}, 
involving highly charged objects.  According to Gauss' theorem, the SCM 
is equivalent to the assumption that the total charge $Ze$ is placed 
at the center of the sphere.  In the ``discrete charge model'' (DCM), 
point charges $(-e)$ are distributed over a sphere of diameter 
$\sigma_d = \alpha \sigma_p$ (with $\alpha < 1$, i.e.\ slightly inside 
the protein surface), in such a way as to minimize the electrostatic 
energy of the distribution.  The resulting charge pattern, well known 
from the classic Thompson problem (see~\cite{Erber} for a recent 
review), is kept fixed throughout.  Such Thompson patterns do not 
correspond to the true charge distribution on any specific protein
(see~\cite{Linse1,Linse2} where  a simple toy model of lysozyme with
different charge ditribution corresponding to solutions of different
pH is constructed)   
but do provide a well defined discrete model for any value of $Z$. 
Note that the discrete distributions are characterized by non-vanishing 
multi-pole moments, depending on the symmetry of the distribution for 
any specific value of $Z$, while the SCM implies vanishing multipoles 
of all orders. 
 
At this stage the SCM and DCM models involve only excluded volume and 
bare Coulomb interactions (reduced by a factor $1/\epsilon$ to account 
for the solvent) between all particles, proteins as well as microions. 
 
The following physical quantities were systematically computed in the 
course of the MD simulations, to be described in the following 
section. 
 
{\bf a)} the density profiles of co and counterions around a single globular 
protein 
\begin{equation} 
\rho_{\pm}(r) = <\sum_{j} \delta(\vec r_j^{\pm} - \vec r)> 
\end{equation}  
Here $\vec r_j^{\pm}$  is the position of the 
$j^{th}$ microion of species $\pm$ relative to the protein center, while 
the angular bracket denotes a canonical average over the microion 
configurations.  For an isolated SCM protein these profiles are 
spherically symmetric, and depend only on the radial distance $r=|\vec 
r|$.  For isolated DCM proteins the profiles are no longer spherically 
symmetric, and may be expanded in spherical harmonics, as discussed in 
the Appendix.  The anisotropy turns out to be weak, and only the 
spherically symmetric component (corresponding to averaging 
$\rho_{\pm}(\vec r)$ over protein orientations) will be shown in the 
following. 
 
{\bf b)} The second quantity, which will be the key input in the 
calculation of $B_2$, is the microion averaged total force $\vec F_1 = 
- \vec F_2$ acting on the center of two proteins, placed at a relative 
position $\vec r= \vec r_1 - \vec r_2$; the force $\vec F_1$ is a 
function of $\vec r$.  Its statistical definition was discussed 
earlier in the context of charged 
colloids\cite{AllahyarovPRL,Trigger,AllahyarovDNA}, and it involves 
three contributions: 
\begin{equation} 
\vec{F}_1=\vec{F}_1^{(1)}+\vec{F}_1^{(2)}+\vec{F}_1^{(3)}. 
\label{force} 
\end{equation} 
$\vec{F}_1^{(1)}$ is the direct Coulomb repulsion between the charge 
distributions on the two proteins; $\vec{F}_1^{(2)}$ is the microion 
induced electrostatic force, while $\vec{F}_1^{(3)}$ is the depletion 
force which may be traced back to the in-balance of the osmotic 
pressure of the microions acting on the opposite sides of protein $1$ 
due to the presence of protein $2$.  $\vec{F}_1^{(3)}$ is directly 
expressible as the integral of the microion contact density over the 
surface of the protein\cite{Atta89,Pias95}. 
 
In the case of the SCM, the microion averaged force depends only on 
the distance $r=|\vec r_{12}|$ between the two proteins.  For the DCM, 
on the other hand, $\vec{F}_1$ is a function of the relative 
orientations of the two proteins, as characterized by the sets of 
Euler angles $\vec \Omega_1$ and $\vec \Omega_2$, i.e.\ 
$\vec{F}_1=\vec{F}_1(\vec r, \vec{\Omega}_1,\vec{\Omega}_2)$. 
 
{\bf c)} Once the force $\vec{F}_1$ has been determined as a function 
of $\vec{r}$, $\vec{\Omega}_1$ and $\vec{\Omega}_2$, one may then 
calculate an orientationally averaged, but distance resolved, effective 
protein-protein pair potential according to 
 
\begin{equation}\label{potential} 
V(r) = \int_r^\infty dr' \langle \frac{\vec{r}'}{|\vec{r}|}\cdot  
\vec{F}_1({\vec r'}, \vec{\Omega}_1, \vec{\Omega}_2)\rangle_{\vec{\Omega}_1,\vec{\Omega}_2}.  
\end{equation} 
where the angular brackets $<...>_{\vec{\Omega}_1 \vec{\Omega}_2}$ 
refer to a canonical statistical average over mutual orientations of 
the two proteins weighted by the Boltzmann factor of the effective 
potential $V_{eff}(\vec{r},\vec{\Omega}_1,\vec{\Omega}_2)$ such that 
$\partial V_{eff}(\vec{r},\vec{\Omega}_1,\vec{\Omega}_2)/\partial 
\vec{r} = - \vec{F}_1(\vec{r},\vec{\Omega}_1,\vec{\Omega}_2)$. 
Explicitly, for any quantity $A(\vec{r},\vec{\Omega}_1,\vec{\Omega}_2)$, 
\begin{equation} 
< A >_{\vec{\Omega}_1,\vec{\Omega}_2} = \frac {\int d 
\vec{\Omega}_1 d\vec{\Omega}_2 A(\vec{r},\vec{\Omega}_1,\vec{\Omega}_2) \exp\{ 
(-V_{eff}({\vec r},\vec{\Omega}_1,\vec{\Omega}_2 )/k_BT)\} } {\int d 
\vec{\Omega}_1 d\vec{\Omega}_2  \exp \{-V_{eff}(\vec{r},\vec{\Omega}_1,\vec{\Omega}_2)/k_BT\} }\label{A} 
\end{equation} 
 
{\bf d)} The second virial coefficient $B_2$ finally follows from the 
expression 
\begin{equation} 
B_2= {1\over 2} \int d \vec{r} \left [ 1-b(r) \right ] 
\end{equation} 
where 
\begin{equation} 
b(r)= ({1\over {8\pi^2}})^2 \int d\vec{\Omega}_1 d\vec{\Omega}_2 
\exp (-V_{eff}(\vec{r},\vec{\Omega}_1,\vec{\Omega}_2)/k_BT). 
\end{equation} 
The angular integrations are trivial in the case of the SCM, where 
$V_{eff}$ depends on $r$.  In the case of the DCM, one may use the 
identity 
\begin{equation} 
b(r)= \exp \left [ -\int_r^{\infty} dr' \ {{d}\over{dr'}} [\ln b(r')] 
  \right ], 
\label{smallb} 
\end{equation} 
to show that $B_2$ may be cast in a form similar to that appropriate 
for the SCM, namely 
\begin{equation} 
B_2 = {1\over 2} \int d\vec{r} \left [1-\exp \{ - V(r)/k_BT\}\right ] 
\label{eqB2} 
\end{equation} 
where $V(r)$ is the potential of the orientationally averaged 
projected force, as defined in Eq.~(\ref{potential}).  As pointed out 
earlier, $B_2$ is directly accessible experimentally by extrapolating 
light scattering data to small wavevectors\cite{protein3} or by taking 
derivatives of osmotic pressure data with respect to 
concentration\cite{Moon,Moon2}.  Results will be presented in the form 
of the reduced second virial coefficient $B_2^* = B_2/B_2^{(HS)}$ 
where $B_2^{(HS)}= 2 \pi \sigma_p^3/3$, i.e. 
\begin{equation}\label{eq2} 
B_2^* = 1 + \frac{3}{\sigma_p^3} \int_{\sigma_p}^{\infty} r^2 dr  
\left[1 - \exp \left\{ - V(r)/k_B T \right\} \right]. 
\end{equation}

\section{Simulation details} 
 
\begin{figure} 
\epsfxsize=12cm 
\epsfysize=10cm  
~\hfill\epsfbox{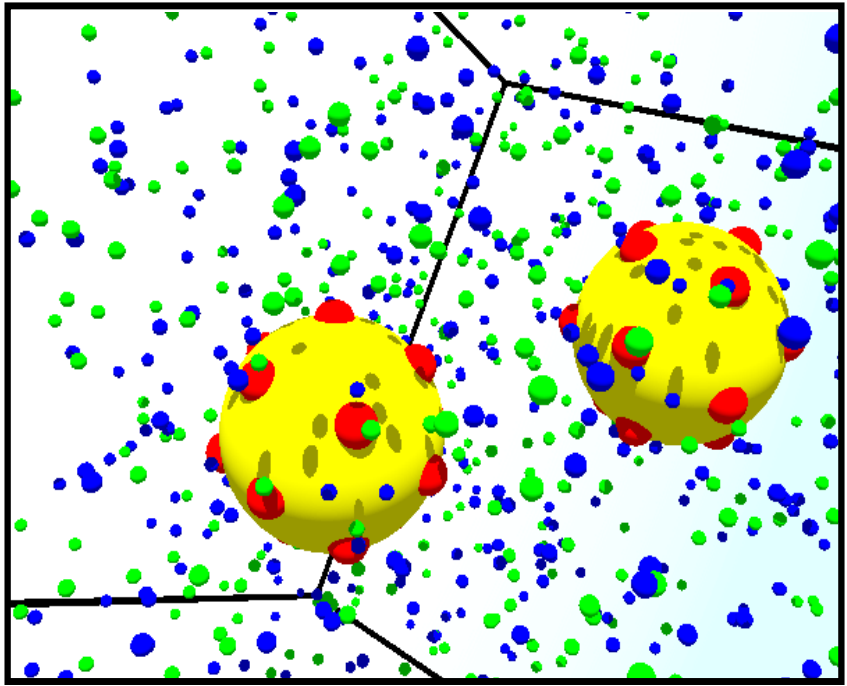}\hfill~ 
\caption{\label{snapshot} 
Snapshot of a typical MD-generated microion configuration around two 
proteins, separated by $r=1.7 \sigma_p$. The proteins carry 15 
discrete charges $-e$ and the monovalent salt density is $C_s = 0.206$ 
Mol/l. The globular protein 
molecules are shown as two large gray spheres. The embedded small dark 
spheres on their surface mimic the discrete protein charges in the DCM 
model. The small gray spheres are counterions, while the black spheres 
are coions.} 
\end{figure}

We study a pair $(N_p=2)$ of spherical proteins with center-to-center 
separation $r$, confined in a cubic box of length $L=4 \sigma_p$,
which also contained monovalent co and counterions in numbers 
determined by their bulk concentrations and overall charge neutrality.
There are $Z N_p$ counterions dissociated from the protein surface,
and added $N_s$ salt ion pairs such that the screening of proteins is
implemented via $N_{+}=N_s$ coions and $N_{-}=N_s + Z N_p$ counterions
in simulation box. A 
snapshot of a typical equilibrium microion configuration around two 
proteins is shown in Figure~\ref{snapshot} for the protein charge 
number $Z = 15$.  The two proteins were placed symmetrically with 
respect to the center along the body diagonal of a cubic simulation 
cell; periodic boundary conditions in three dimensions were adopted.  $L$ 
was chosen such that the box length is much larger than the range of 
the total (effective) protein-protein interaction, so that the results 
are independent of $L$ for non-zero salt concentration. The long-range 
electrostatic interactions between two charged particles in the 
simulation box with periodic boundary conditions were modified using 
the Lekner summation method of images \cite{lekner}. For our model 
to be a rough representation of lysozyme, we chose $\sigma_p = 4 nm$, 
and three different protein charges $Z = 6, 10$ and $15$, 
corresponding to three different values of the solution pH. The 
microion diameter is fixed to be $\sigma_c= 
\sigma_p/15 = 0.267 
nm$.

 For both the SCM and the DCM, the contact coupling parameter between 
a protein and a microion, namely $\Gamma = 2 e^2 /[\epsilon k_B T( 
\sigma_p-\sigma_d + \sigma_c)]$ for the DCM, and $\Gamma = 2 
Ze^2/[\epsilon k_B T(\sigma_p + \sigma_c)]$ for the SCM, are 
comparable, and of the order of $\Gamma\approx3$ at room temperature. 
We fixed the dielectric constant of water to be $\epsilon=81$ and the 
system temperature to be $T=298 K$. Varying salt concentration for 
fixed protein charge $Z$ corresponds to a fixed solution 
pH\cite{Tanford}. 
 
Details of the runs corresponding to different salt concentrations are 
summarized in Table I. Note that the Debye screening length $r_D$, 
defined by 
\begin{equation} 
r_D=\sqrt 
{\frac {\epsilon k_B T V^{'}} {8 \pi (N_s+Z) q_s e^2} }, 
\label{debye} 
\end{equation} 
is less than $10\AA$ for salt concentration beyond 0.1M\@. 
Here 
\begin{equation} 
V^{'}=V-\frac{\pi N_p}{6} (\sigma_p^3+\sigma_c^3)  , 
\label{volume} 
\end{equation} 
is the accessible volume for salt ions such that the salt
concentration $C_s$ is $N_s/V^{'}$. Thus, the point charges on the protein 
surface are effectively screened from each other \cite{Broide}.  For 
each of the runs indicated in Table I, the distance-resolved effective 
forces and interaction potentials are calculated according to 
Eqs.~(\ref{force},\ref{potential}). The statistical averages over 
microion configurations leading to $\vec{F}_1^{(2)}$ and 
$\vec{F}_1^{(3)}$ were evaluated from time averages in the Molecular 
Dynamics (MD) simulations. 
 
\section{Microion distributions around a single protein} 
\begin{figure} 
   \epsfxsize=12cm 
   \epsfysize=12cm 
   ~\hfill\epsfbox{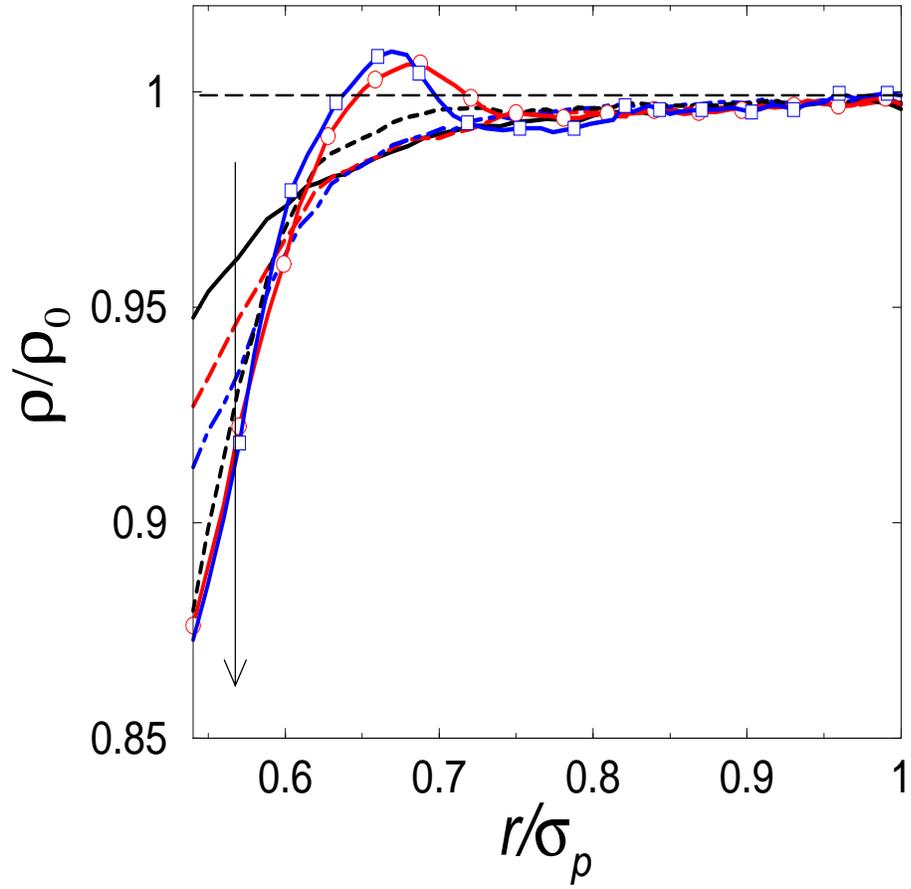}\hfill~  
\caption{Normalized 
   total salt density profiles $\rho(r)$ near single {\it neutral 
   sphere}. $\rho_0 = N_s/V^{'}$ is bulk density. The added salt 
   concentration is increased from top to bottom (see the arrow which 
   refers to the $\rho(r)$ near protein surface) according to runs 1-5, 7, 
   9. } 
   \label{neutralsphere} 
\end{figure}

First, as a reference, consider a single protein ($N_p=1$) placed at 
the center of the simulation box. We calculated spherically averaged, 
radial microion density profiles $\rho(r) = 
\rho_+(r) + \rho_-(r)$ in the immediate vicinity of the 
protein surface.  For a single {\it neutral} sphere in a salted 
solution, results for $\rho(r)$ are drawn in 
Figure~\ref{neutralsphere}. There is a marked depletion in the 
microion density, signaled by a minimum of $\rho(r)$, well below the 
asymptotic bulk value.  The depletion is enhanced upon increasing the 
salt concentration.  At sufficiently high salt concentrations, this 
minimum is followed by a weak, but detectable, ion layer (see 
corresponding lines for runs 7 and 9 in 
Figure~\ref{neutralsphere}). The formation of a depletion zone is {\em 
not} a consequence of the direct (hard core) interaction between salt 
ions and the protein surface, since the position of the observed layer 
is significantly further away from the protein surface than one ion 
diameter.  A rough estimate for the distance between layer and neutral 
sphere gives a value of $2.5\sigma_s$, or equivalently $0.17\sigma_p$. 
For runs 7 and 9, where the ion layer emerges, this distance is 
of the order of an average ion separation $r_s$ in the system and
twice the Debye screening length $r_D$ as well (see Table I). Obviously, 
it is the small ion correlations which lead to the peak formation in 
the salt density profiles. As an intuitive argument, we posit that the 
lack of mutual polarization in dense salt solution near neutral 
surfaces causes the ion depletion. Qualitatively similar depleted 
density profiles were observed in Lennard-Jones system confined 
between neutral planes~\cite{Berard} and in Yukawa 
mixtures~\cite{Ardmixture}. Furthermore, an effective force that pushes a
single ion toward regions of higher salinity is predicted 
 within Debye-H\"{u}ckel theory for interfacial geometries \cite{netz}.

\begin{figure} 
   \epsfxsize=10cm 
 \epsfysize=11cm 
 \epsfbox{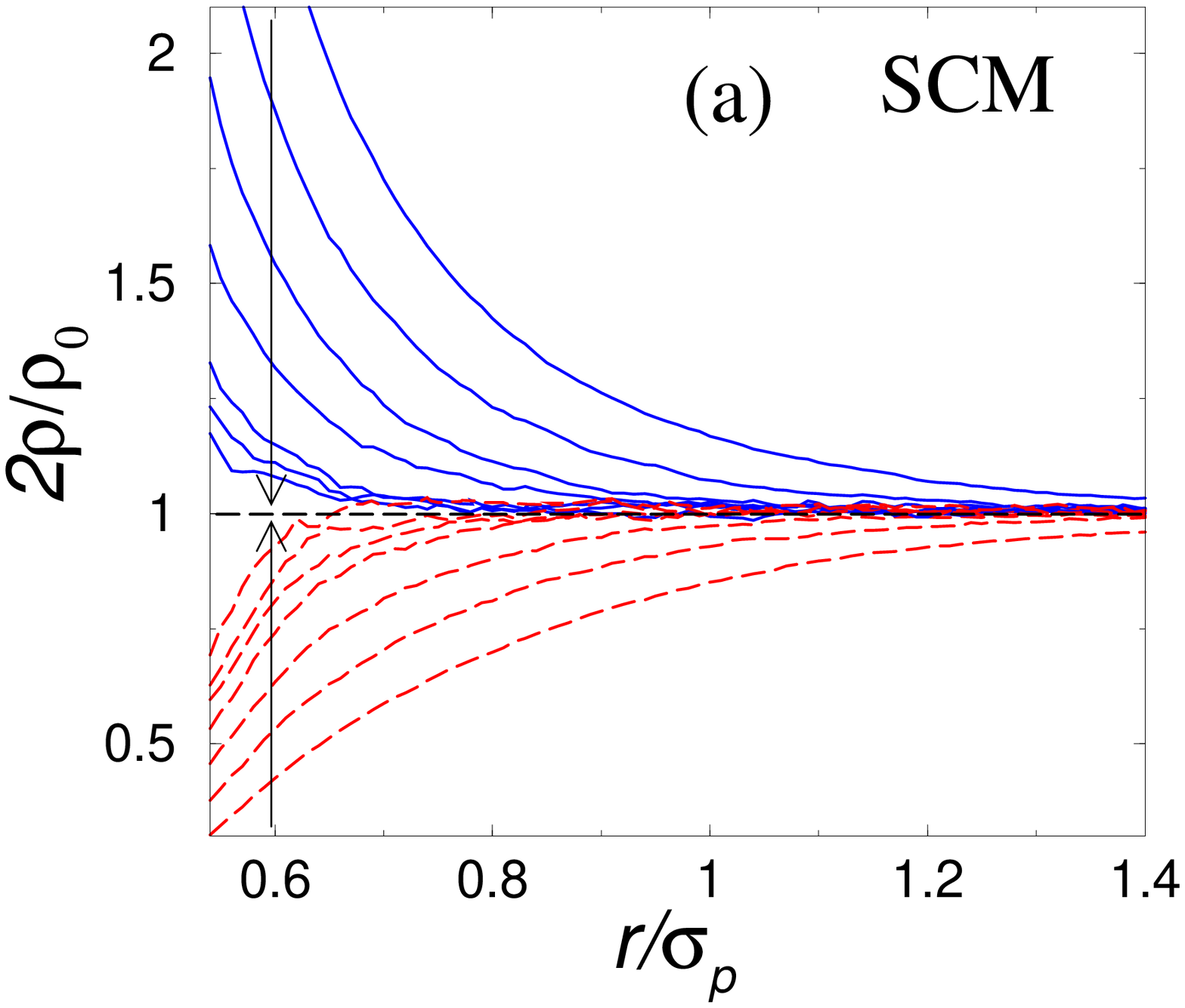}   \vskip -2.5cm 
\epsfxsize=10cm 
\epsfysize=11cm 
 \epsfbox{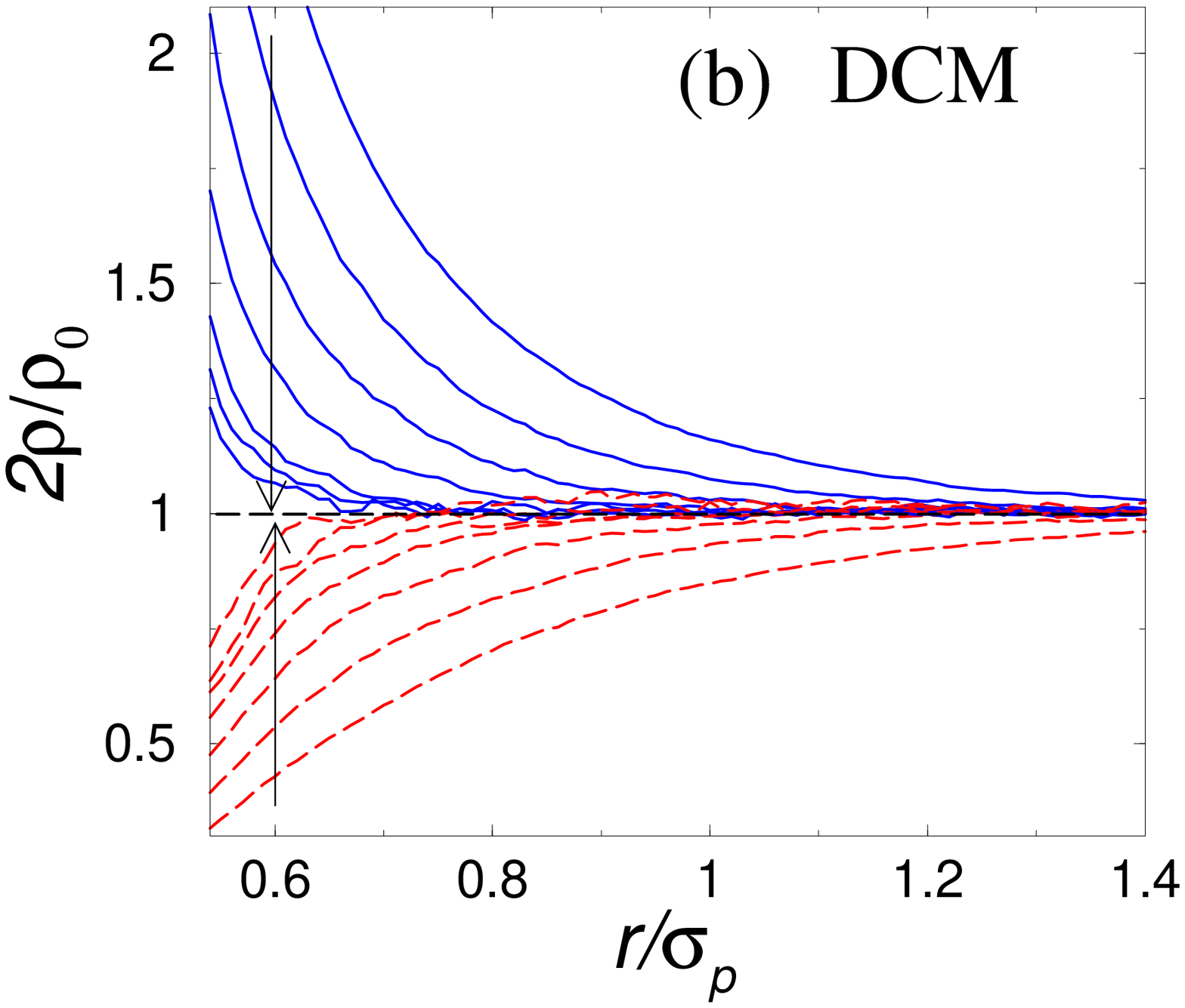} 
   \caption{Rescaled density profiles of small ions near single 
   protein surface for the SCM (a) and DCM (b) models. The protein 
   charge is $Z=10$. The added salt concentration is increased (shown 
   by an arrow) from top to bottom for solid lines (counterions) and 
   from bottom to top for dashed lines (coions), according to runs 
   2-5, 7, 9, 11.}
  \label{coandcounterions} 
\end{figure}

\begin{figure} 
   \epsfxsize=10cm 
   \epsfysize=11cm 
~\hfill\epsfbox{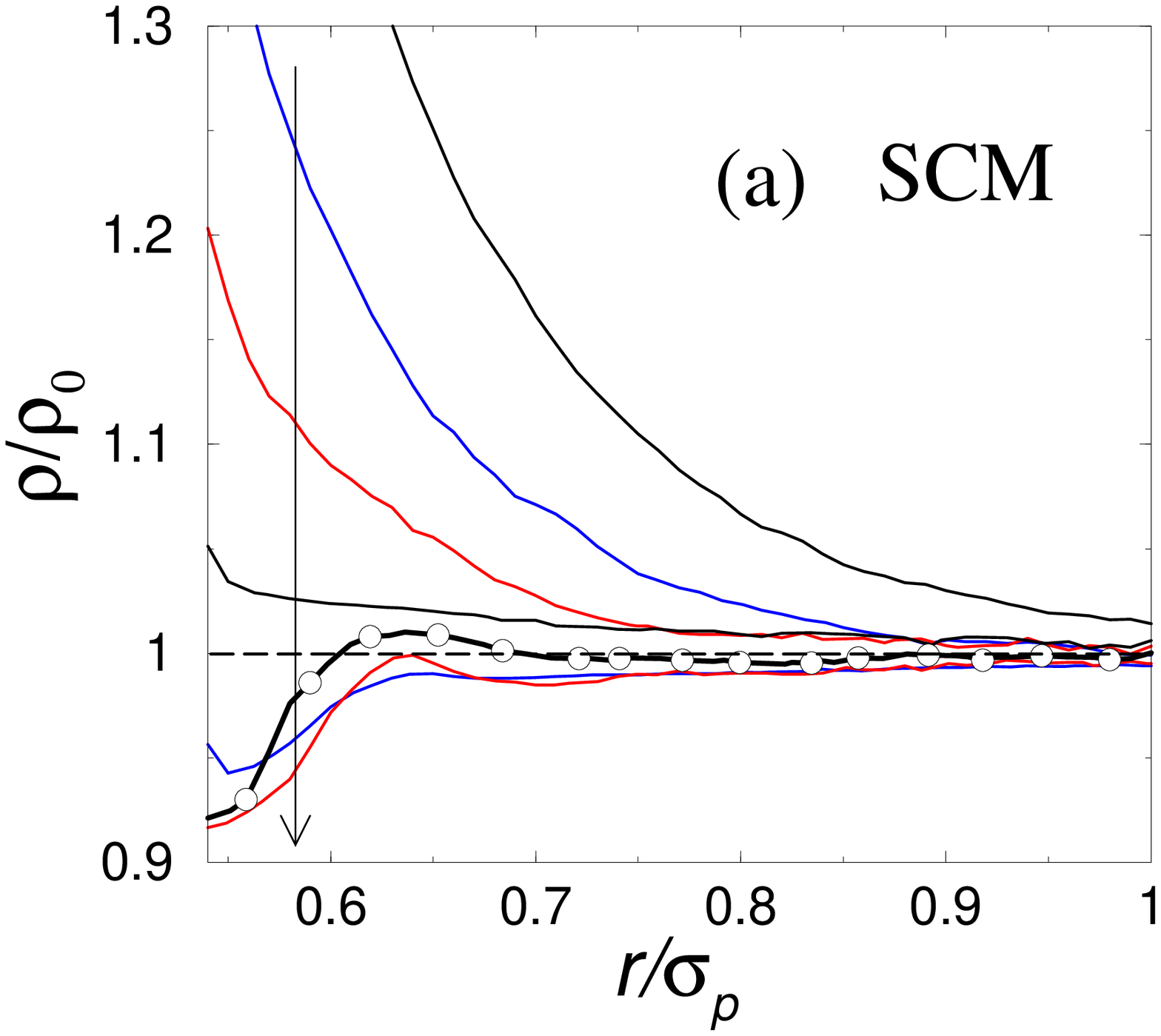}\hfill~   \vskip -2.5cm 
   \epsfxsize=10cm 
   \epsfysize=11cm 
~\hfill\epsfbox{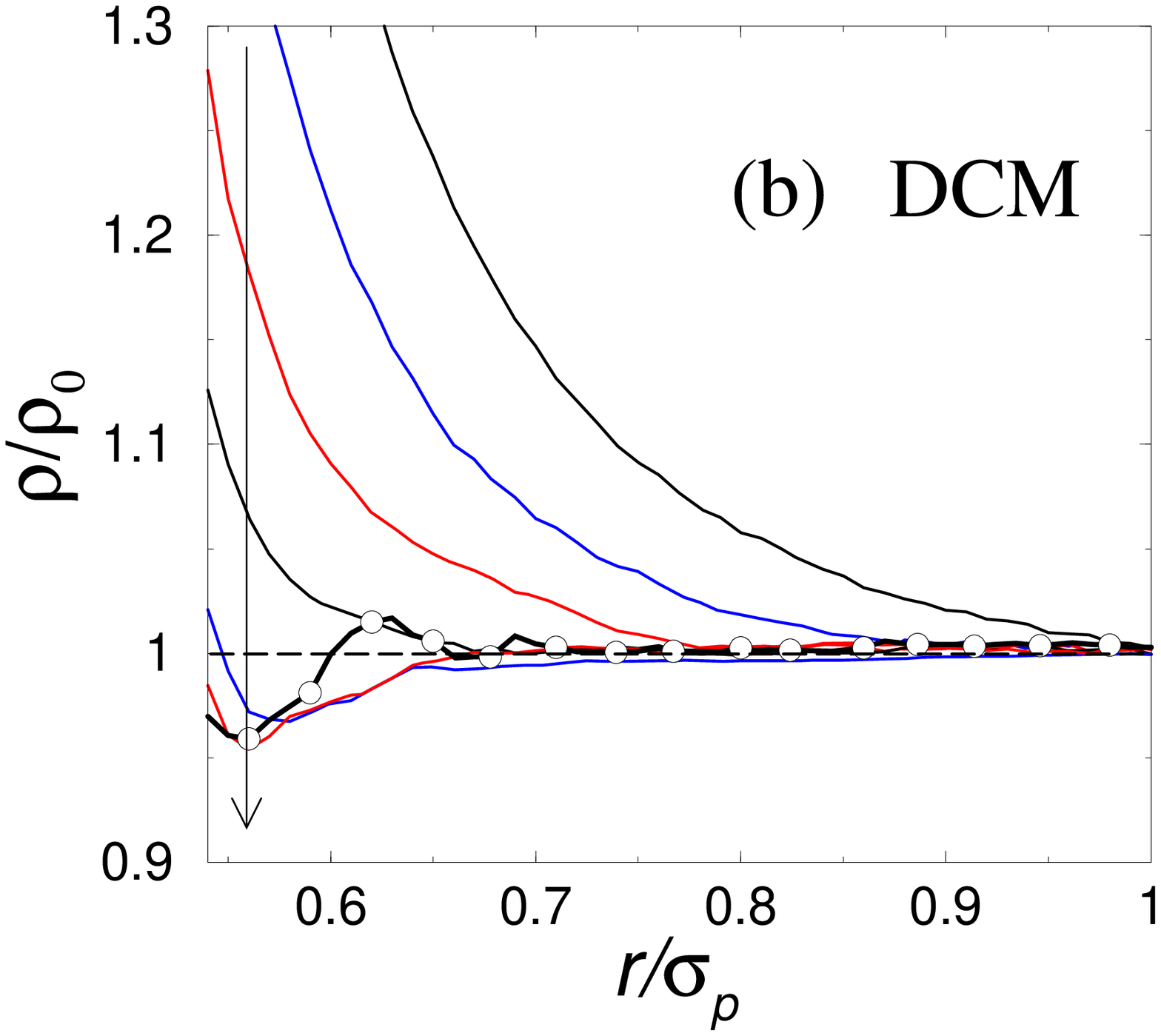}\hfill~  
   \caption{Same runs as Figure~\ref{coandcounterions}, but now for the 
   total salt density near single protein surface. The arrow (a
   direction of added salt increase) applies to all runs except
   run 11, which is shown as a solid line with symbols. } 
     \label{z10totaldensity} 
\end{figure}

\begin{figure} 
   \epsfxsize=11cm 
   \epsfysize=10.cm 
  \hspace {-0.9cm} ~\hfill\epsfbox{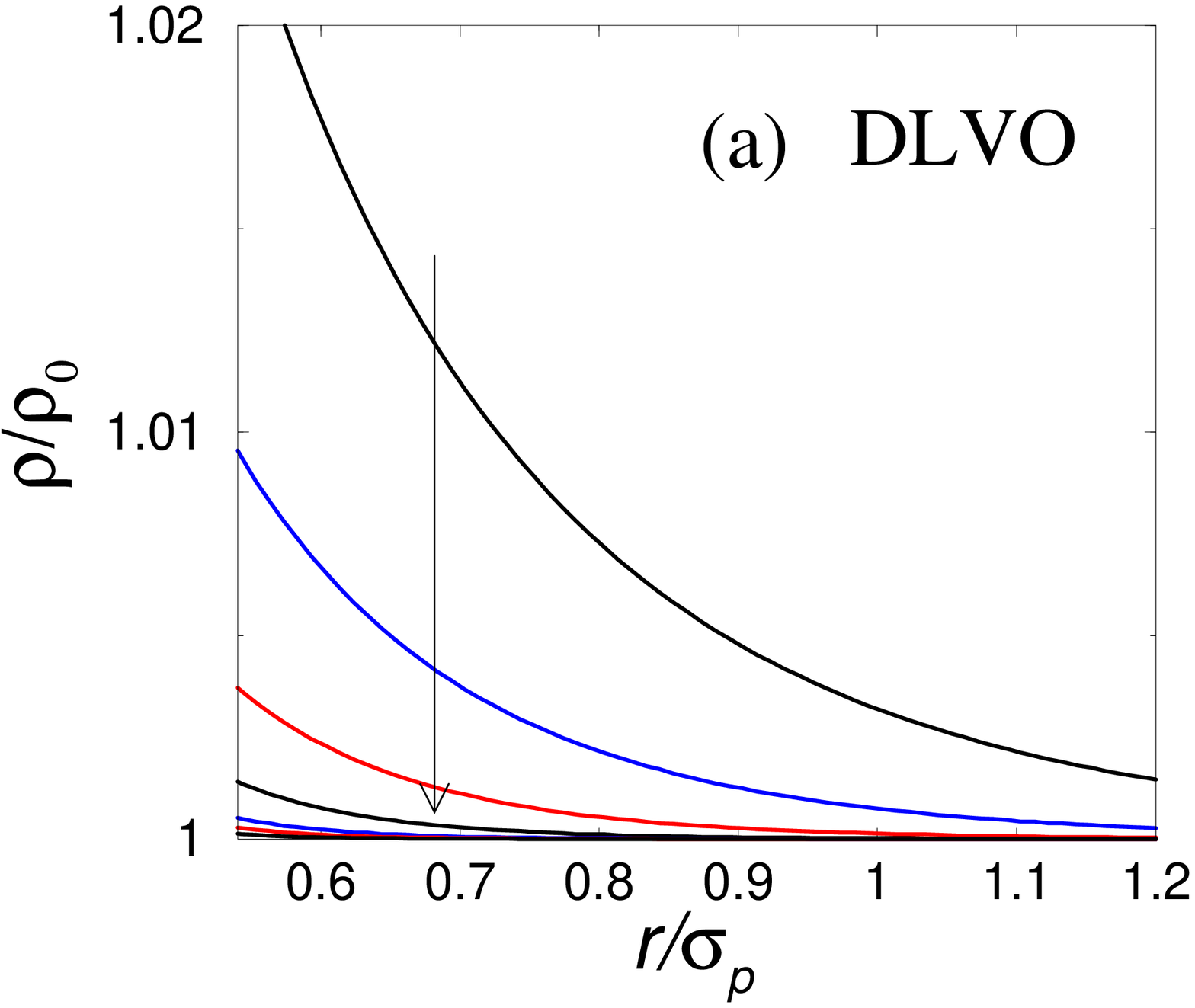}\hfill~   \vskip -2.5cm 
   \epsfxsize=10cm 
   \epsfysize=11cm 
~\hfill\epsfbox{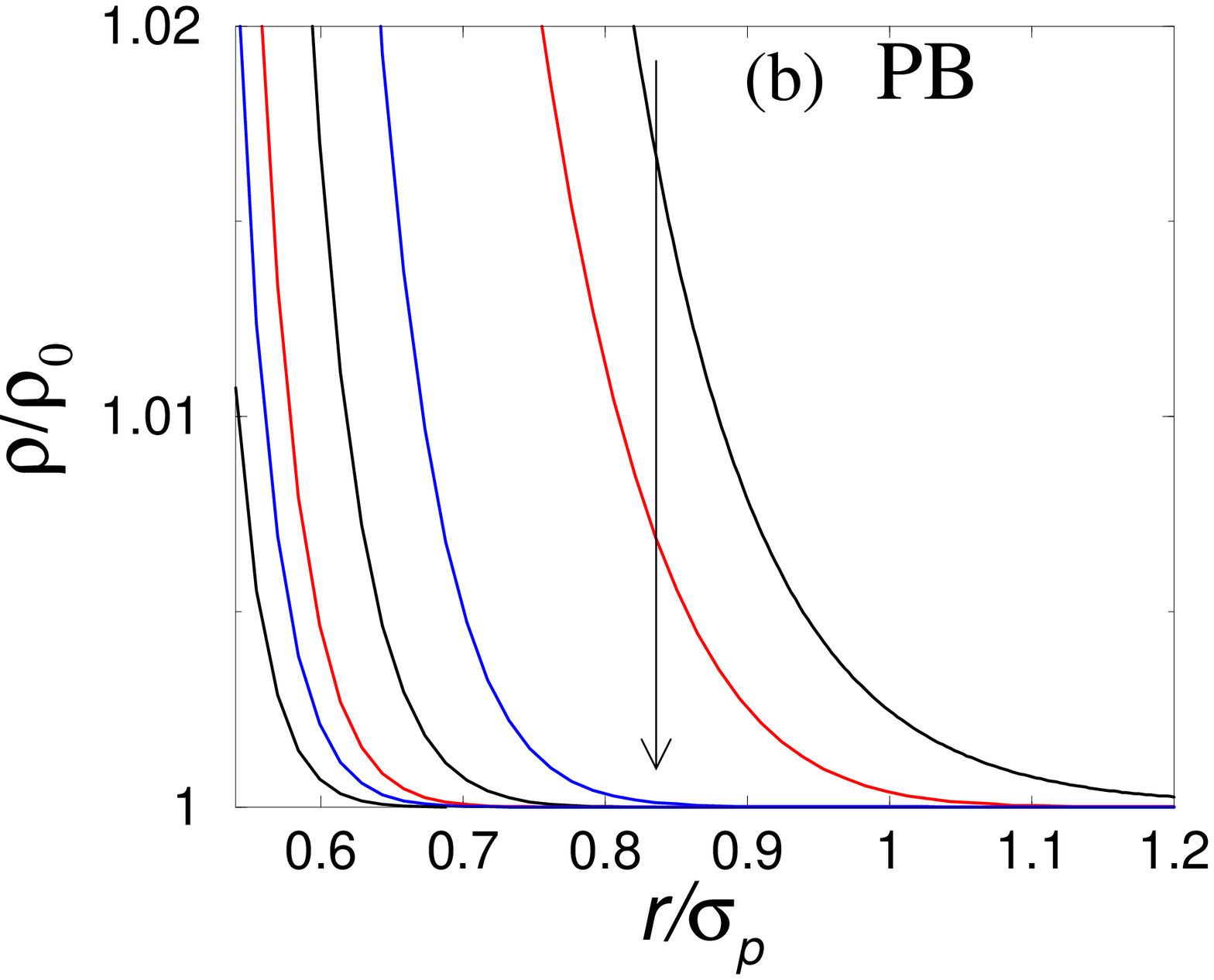}\hfill~ 
   \caption{Same runs as in Figure~\ref{z10totaldensity}, but now for 
   DLVO theory (a), and non-linear PB theory (b); both are for the 
   SCM.}  \label{adensityDLVO} 
\end{figure}

Next we consider a protein sphere with charge number $Z=10$. The 
density profiles of small ions are shown in 
Figure~\ref{coandcounterions} for both the SCM and 
DCM. Figure~\ref{z10totaldensity} represents the corresponding total 
salt densities, as a sum of co- and counterion densities from 
Figure~\ref{coandcounterions}. At the lower salt concentration (up to 
run 5) the SCM and DCM models both yield an accumulation of the 
microion density near ion-protein contact, in semi-quantitative 
agreement with the prediction of standard Poisson-Boltzmann (PB) 
theory. For rising ionic strength the total microion density gets 
depleted near the protein surfaces, as in the previously considered 
case of a neutral sphere. Remarkably, this depletion occurs both with 
the SCM and DCM and contradicts the PB prediction.  The intuitive 
picture is that a microscopic layer of counterions is formed around 
the proteins. An additional salt pair now profits more from the bulk 
polarization than from the protein surface polarization and is thus 
excluded from this layer. By normalizing the profiles to the total 
bulk density, this effect becomes visible as a depletion zone in 
Figure~\ref{z10totaldensity}, where a noticeable difference between 
the SCM and DCM profiles also emerges. Whereas the DCM predicts a 
contact value $\rho_c(r=(\sigma_p + \sigma_c)/2)$ larger than the bulk 
value, SCM predicts a much stronger microion depletion near 
contact. Together with this, the contact value of the DCM model is
always larger than that of the SCM model for the same salt
concentrations. This finding illustrates the sensitivity of correlation 
effects to the assumed charge pattern at the surface of a protein. 
This correlation effect is, of course, absent in the (non-linear) PB 
and DLVO theories, which always predict a monotonically decreasing 
density profile $\rho(r)$ (see Figure~\ref{adensityDLVO}).  Within 
DLVO theory the density of plus and minus salt ions near protein 
surface in the SCM model are defined as 
\begin{eqnarray} 
\rho_{+}(r) = - \frac {Z_{DLVO}}{q_s} \frac {k_{+}^2} {4 \pi}  
\frac {e^{-k_D r}}{r} 
 + \frac           {   2 \overline{\rho}_{+}    \overline{\rho}_{-}  }  
                   {     \overline{\rho}_{+}  + \overline{\rho}_{-}  }, 
\nonumber                   \\ 
\rho_{-}(r) =  \frac {Z_{DLVO}}{q_s} \frac {k_{+}^2} {4 \pi}  
\frac {e^{-k_D r}}{r} 
 + \frac           {   2 \overline{\rho}_{+}    \overline{\rho}_{-}  }  
                   {     \overline{\rho}_{+}  + \overline{\rho}_{-}  }. 
\end{eqnarray} 
where 
\begin{eqnarray} 
\overline{\rho}_{+} = \frac {N_s + 2Z} {V^{'}}, ~~~\overline{\rho}_{-} 
= \frac {N_s} {V^{'}}, ~~ 
Z_{DLVO} = Z \frac {\exp (k_D \sigma_p /2) } { 1 + k_D \sigma_p /2}, 
\nonumber                   \\ 
k_D^2 = k_{+}^2 + k_{-}^2, ~~~~~~k_{-}^2 = 4 \pi q_s^2 e^2 
\overline{\rho}_{-} /\epsilon k_B T ~,~~~~~ 
 k_{+}^2 = 4 \pi q_s^2 e^2 \overline{\rho}_{+} /\epsilon k_B T~. 
\end{eqnarray} 
Nonlinear PB equations were solved via iterations of the potential of 
a homogeneously charged sphere placed at the center of a spherical 
cell. The cell radius was determined by the given protein 
concentration. 
 
\begin{figure} 
   \epsfxsize=11cm  
   \epsfysize=12cm 
~\hfill\epsfbox{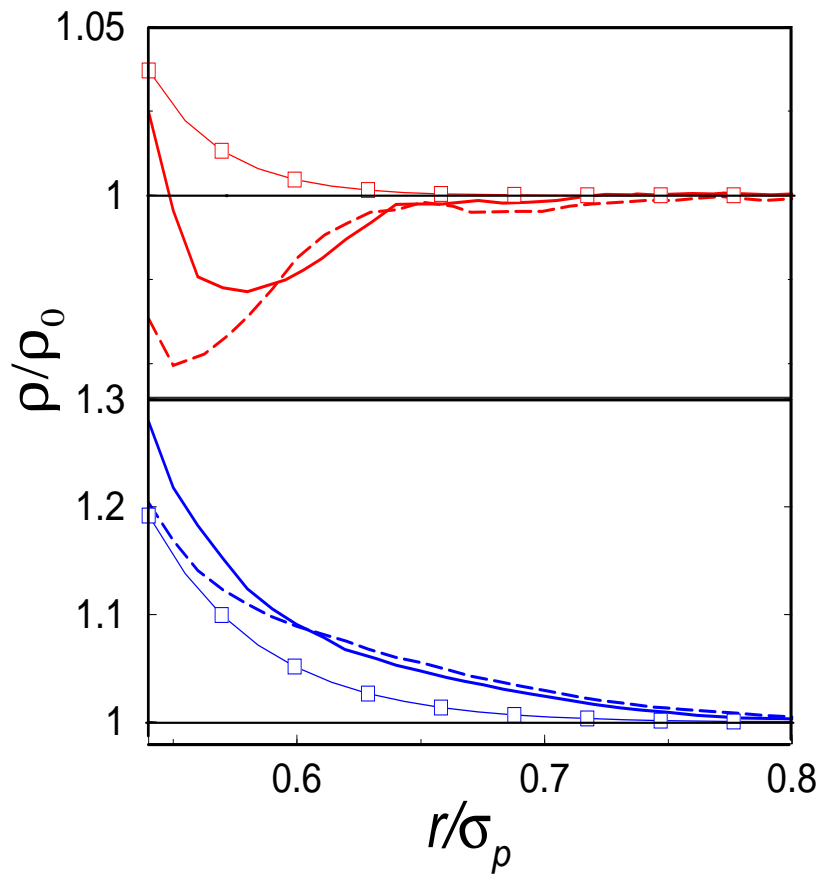}\hfill~ 
\caption{Total density profiles $\rho(r)$ 
of salt ions around a single protein with $Z=10$, for run 4 (bottom 
set of curves) and run 7 (upper set of curves), comparing DCM 
simulations (solid line), SCM simulations (dashed line), and nonlinear 
Poisson-Boltzmann theory (squares connected by lines).} 
\label{dcmscmpb} 
\end{figure}

A direct comparison between SCM, DCM models and non-linear PB theory 
is shown in Fig.~\ref{dcmscmpb} for two of the higher salt 
concentrations from Fig.~\ref{z10totaldensity}.  For the  intermediate salt 
concentration $C_s=0.206$ Mol/l (run 4) both simulation and theory 
predict a monotonic decrease of salt density away from the protein 
surface, whereas, in the case of a dense salt, $C_s=0.824$ Mol/l (run 
7), simulation results strongly deviate from the PB prediction. 
 
A multipole expansion of the total salt number density in the DCM, 
discussed in Appendix A, demonstrates that the higher order expansion 
coefficients are strongly damped and much weaker than the 
zero-order homogeneous term shown in
Figs. \ref{coandcounterions}, \ref{z10totaldensity} and \ref{dcmscmpb}. 
 
\subsection{Effective force and $B_2$ for a protein pair} 
 
\begin{figure} 
   \epsfxsize=9cm 
   \epsfysize=8.5cm
  \hspace {0.7cm} \epsfbox{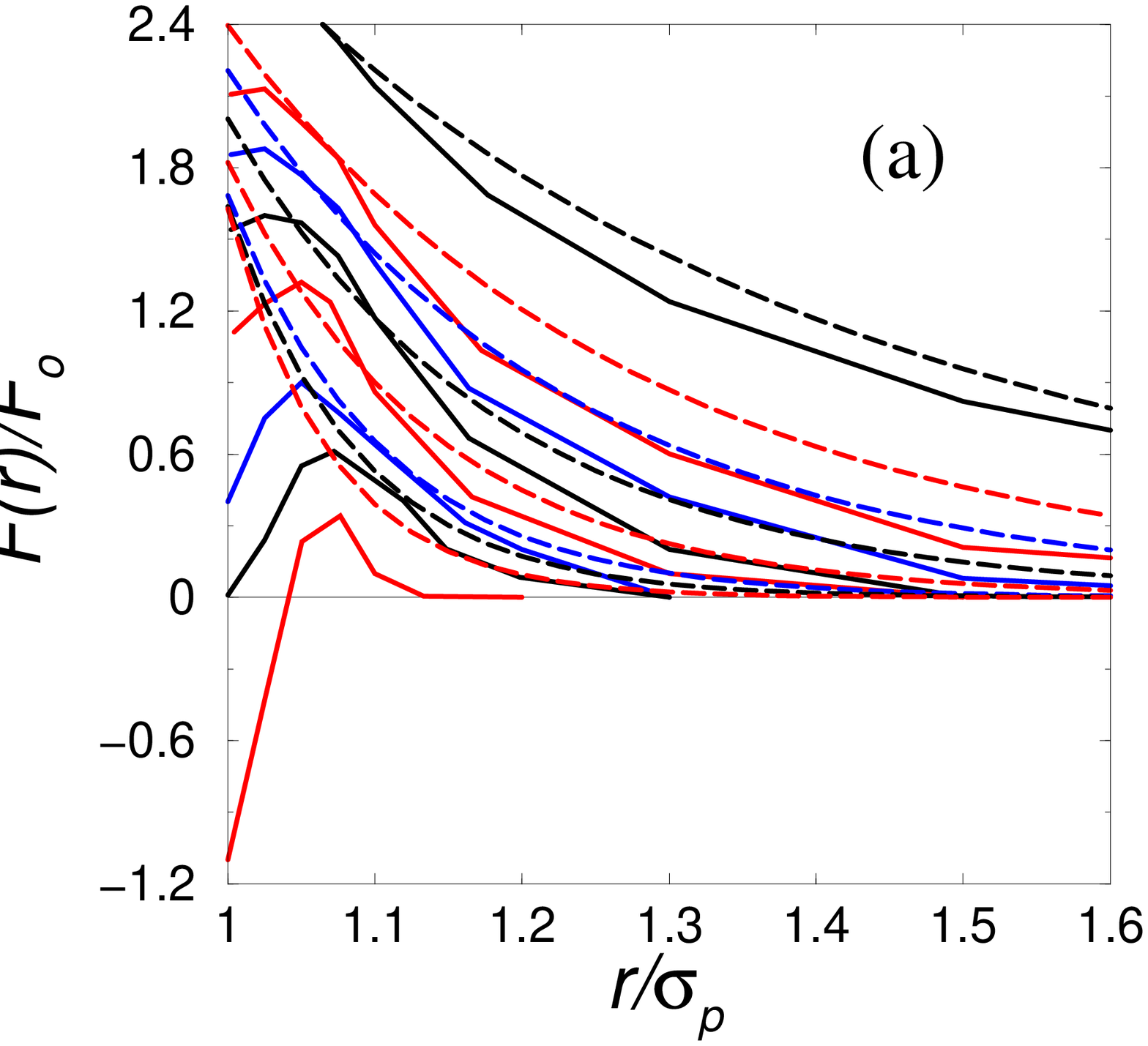}     \vskip 1cm
   \epsfxsize=9.7cm 
   \epsfysize=8.5cm 
   \epsfbox{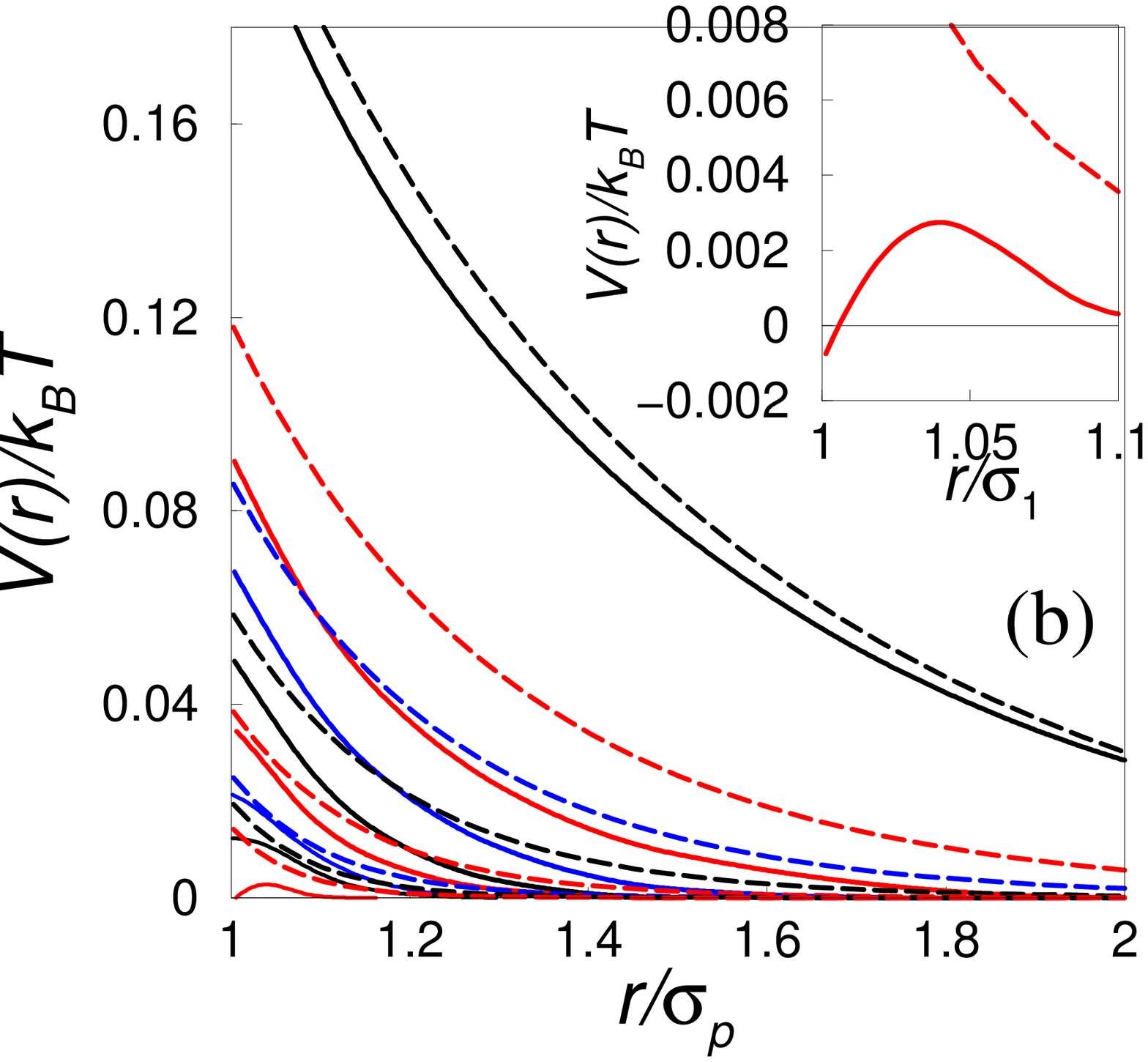} 
   \caption{Total interaction force $F(r)$ (a) and interaction 
   potential $V(r)$ (b) versus dimensionless separation distance 
   $r/\sigma_p$ within the SCM for a protein charge $Z=10$. The force 
   is divided by $F_0=k_BT/ \lambda_B$, where $\lambda_B=e^2/\epsilon 
   k_B T$ is the Bjerrum length. The added salt concentration is 
   increased from top to bottom, according to runs 1-5, 7, 9, 
   11. Dashed lines correspond to the DLVO model. The inset in (b) 
   shows in more detail the differences between the SCM simulations 
   and the DLVO model potential for run 11.} 
   \label{scmforcepotential} 
\end{figure}

\begin{figure} 
   \epsfxsize=10cm 
   \epsfysize=10cm 
  \epsfbox{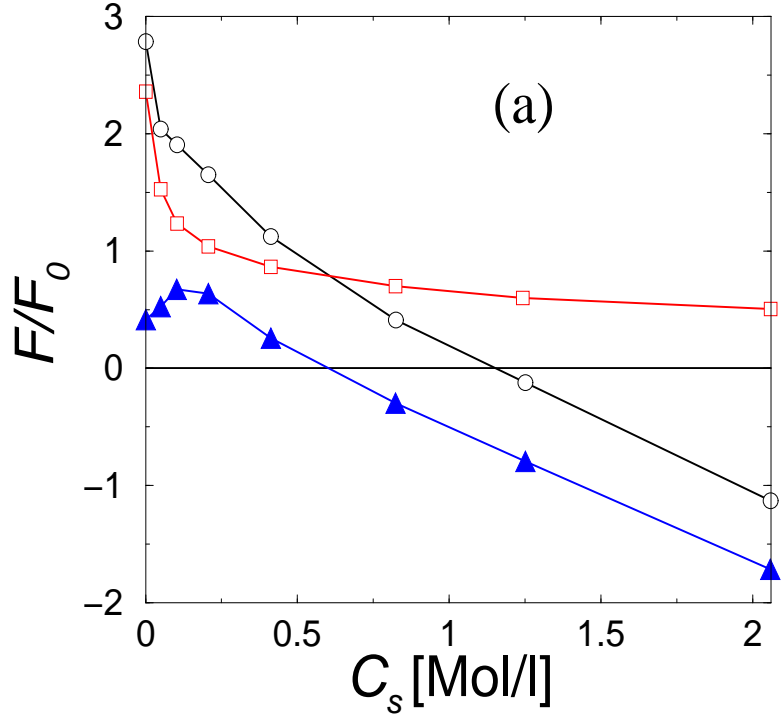} 
   \epsfxsize=10cm 
   \epsfysize=10cm 
   \epsfbox{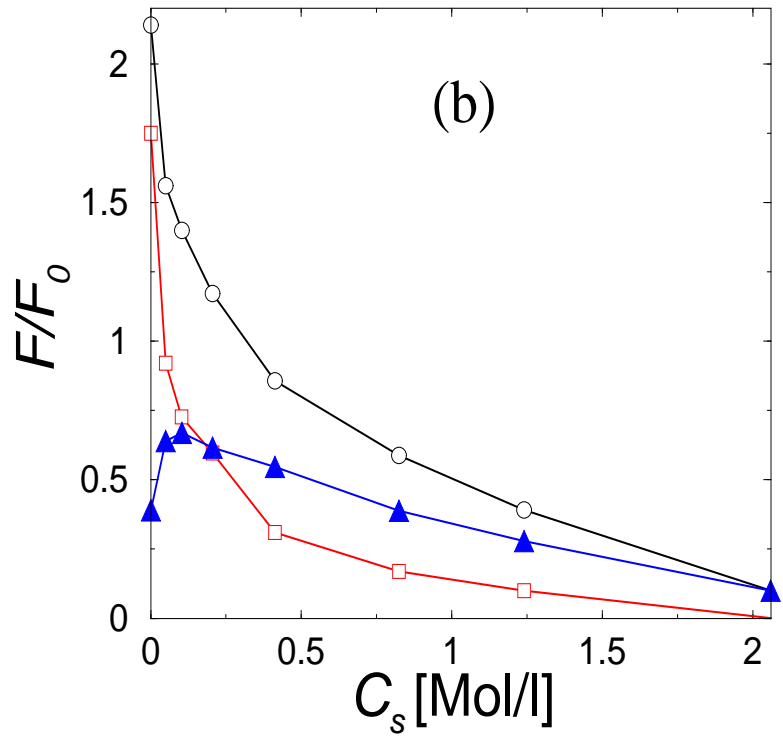} 
   \caption{The total interaction force $F$(circles) and its 
   electrostatic, $F^{(2)}$ (squares) and entropic $F^{(3)}$ 
   (triangles) components versus salt concentration. The separation 
   distance is fixed at (a) $r/\sigma_p=1$ and (b) 
   $r/\sigma_p=1.1$. The simulations are for the SCM with $Z=10$, and 
   show that at high salt concentrations, the entropic force 
   dominates.}  \label{scmforcecomponents} 
\end{figure}

Next we calculate the angularly averaged effective interaction force 
$F(r)=-\frac {dV(r)}{dr}$ and potential $V(r)$ between two proteins embedded in a sea of 
small salt ions.  Simulation results for the simpler case of the SCM, are 
plotted in Figure~\ref{scmforcepotential} for $Z=10$ and compared to 
the DLVO predictions.  There is a systematic deviation between the 
theoretical and simulation results.  While the DLVO 
theory~\cite{Verwey} potential 
\begin{equation} 
{U}^{(DLVO)}(r)= \frac{Z_{DLVO}^2 e^2}{\epsilon r} \exp(- r/r_D ), 
\label{FDLVO} 
\end{equation} 
always results in repulsive forces, simulations indicate the
possibility of 
an attraction between proteins for large salt concentrations. The 
force $F(r)$ at the higher salt concentrations $C_s$, 
shows a maximum at a distance $r$ nearly equal to the ion 
diameter. Note that, for the highest salt concentration considered, 
$C_s=2.061$ Mol/l (run 11), where the electrostatic interactions are 
almost completely screened out, the effective force $F(r)$ is 
dominated by entropic effects; it is reminiscent of entropic depletion 
force of hard sphere system. The corresponding potential is negative 
at short distances as shown in the inset of 
Figure~\ref{scmforcepotential}, and is related to the depletion in the 
microion total density profiles $\rho (r)$ around an isolated protein, 
shown in Figure~\ref{z10totaldensity}. We note that such an entropic 
attraction is not contained in DLVO theory.  Its origin is also 
different from the salting out effect studied in 
\cite{Rose96,zamorazukoski,howardtwigg,rosenberger1993,cacioppopusey} 
or the macroion overcharging effect studied in\cite{Messina}.  In 
Figure~\ref{scmforcecomponents} the salt dependence of the total 
interaction force $F(r)$ (Eq.(\ref{force})) is broken down into its 
components $F^{(2)}$ and $F^{(3)}$ for two values of $r$.  This helps
to show that at large salt concentrations, it is indeed the entropic 
component that causes the force to be attractive for run 11 in 
Figs~\ref{scmforcepotential} and ~\ref{scmforcecomponents}a.  Finally, 
we mention that the range of attraction observed here will depend on 
the electrolyte (salt ion) size.  This feature of our model may hint 
at a cause for the salt specificity observed in salting out 
experiments on protein crystallization\cite{protein2}.  In addition, 
as shown in \cite{Prausnitz1}, an entropic attraction from the 
electrolyte could lead even to phase separation in colloid-electrolyte 
mixtures.  Including this term leads to an effective Hamaker constant 
-- describing the dispersion interactions -- that is lower, and in 
better accord with experimental findings \cite{Tardieu,Prausnitz2}. 
 
 The same calculations were carried out for the other two protein 
charges for the SCM model, $Z=6$ and $Z=15$, with qualitatively 
similar results to those obtained for $Z=10$. 
 
\begin{figure} 
   \epsfxsize=9cm 
   \epsfysize=9cm 
\epsfbox{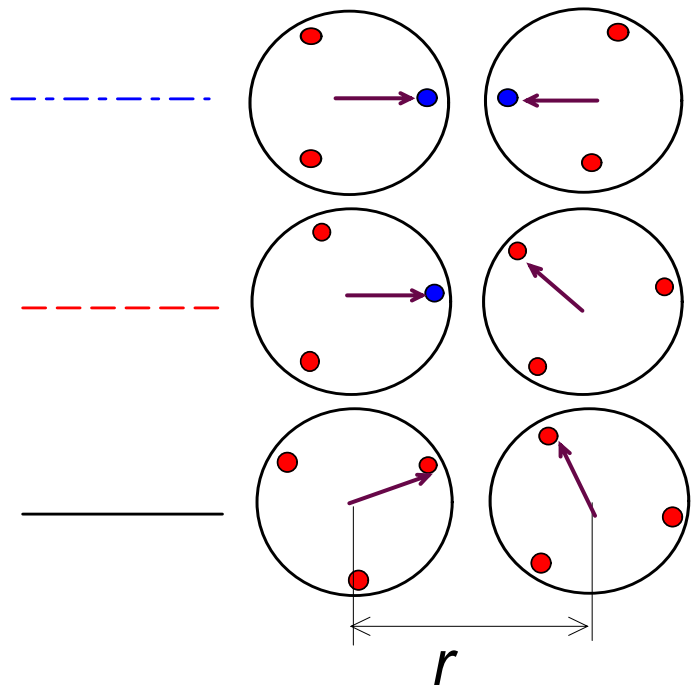} \vskip 1cm
   \epsfxsize=10cm 
   \epsfysize=10cm 
\epsfbox{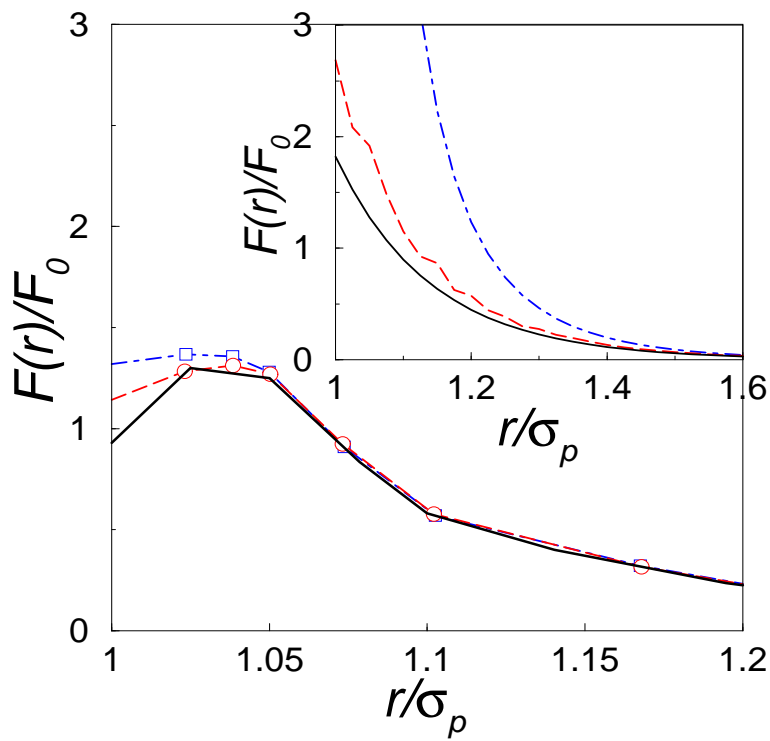} 
   \caption{(a) An illustration of three different mutual 
     orientations of two proteins. 
Points inside spheres represent protein charges in the DCM\@.  
 (b) Total interaction force $F(r)$ versus dimensionless separation 
     distance $r/ \sigma_p $ for mutual orientations shown in (a) for 
     run 5 and $Z=10$ in the DCM. The inset shows the same, but for a Yukawa 
     segment model.} 
     \label{dcmomega} 
\end{figure}

\begin{figure} 
   \epsfxsize=10cm 
   \epsfysize=10cm 
   \epsfbox{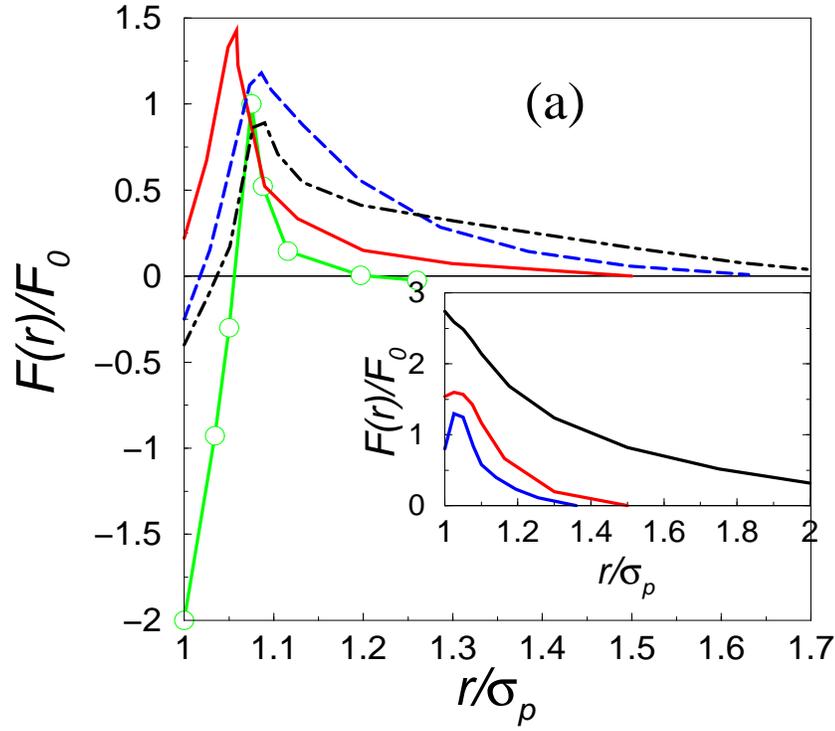} 
   \epsfxsize=10cm 
   \epsfysize=10cm 
   \epsfbox{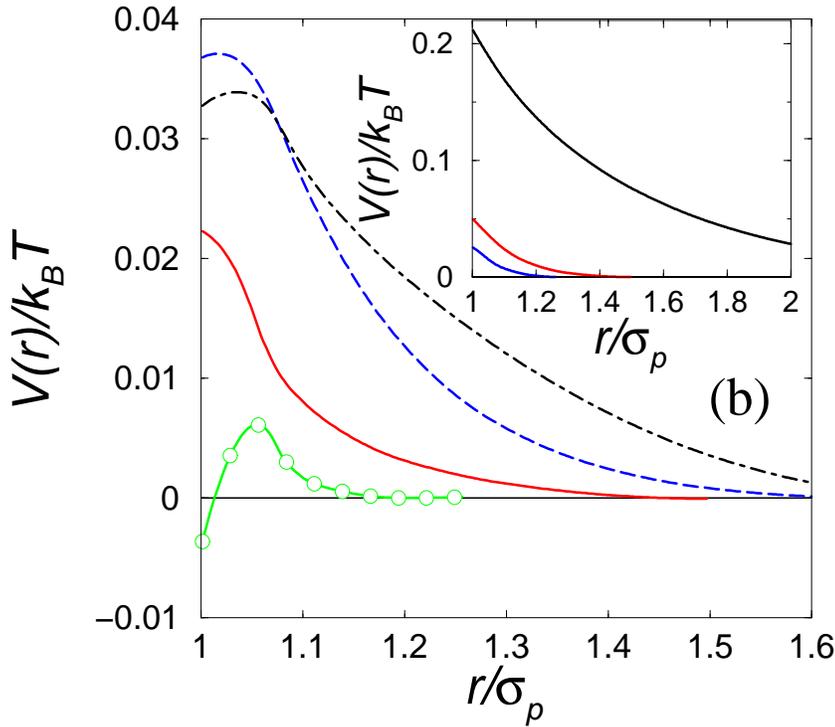} 
   \caption{Total interaction force $F(r)$ (a) and interaction 
   potential $V(r)$ (b) versus dimensionless separation distance 
   $r/\sigma_p$ for the DCM at $Z=10$.  Solid line- run 7, dashed 
   line- run 8, dot-dashed line- run 9, open circles- run 11.
 The inset shows low 
   salt concentrations, from top to bottom: runs 1, 4, 5.  
}  \label{dcmforcepotentialz10} 
\end{figure}

\begin{figure} 
   \epsfxsize=10cm 
   \epsfysize=10cm 
   \epsfbox{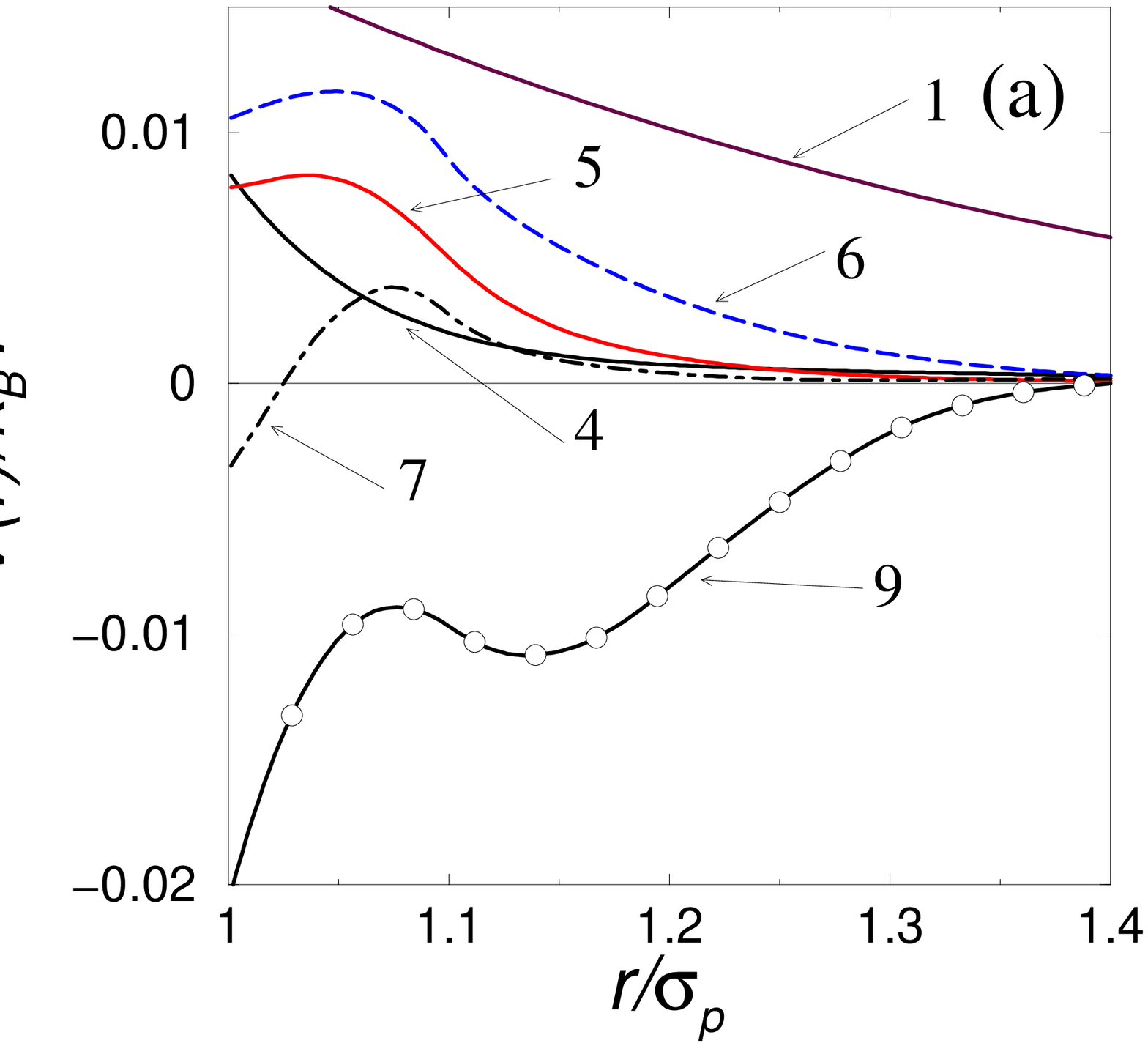} 
   \epsfxsize=10cm 
   \epsfysize=10cm 
\hspace {0.3cm} \epsfbox{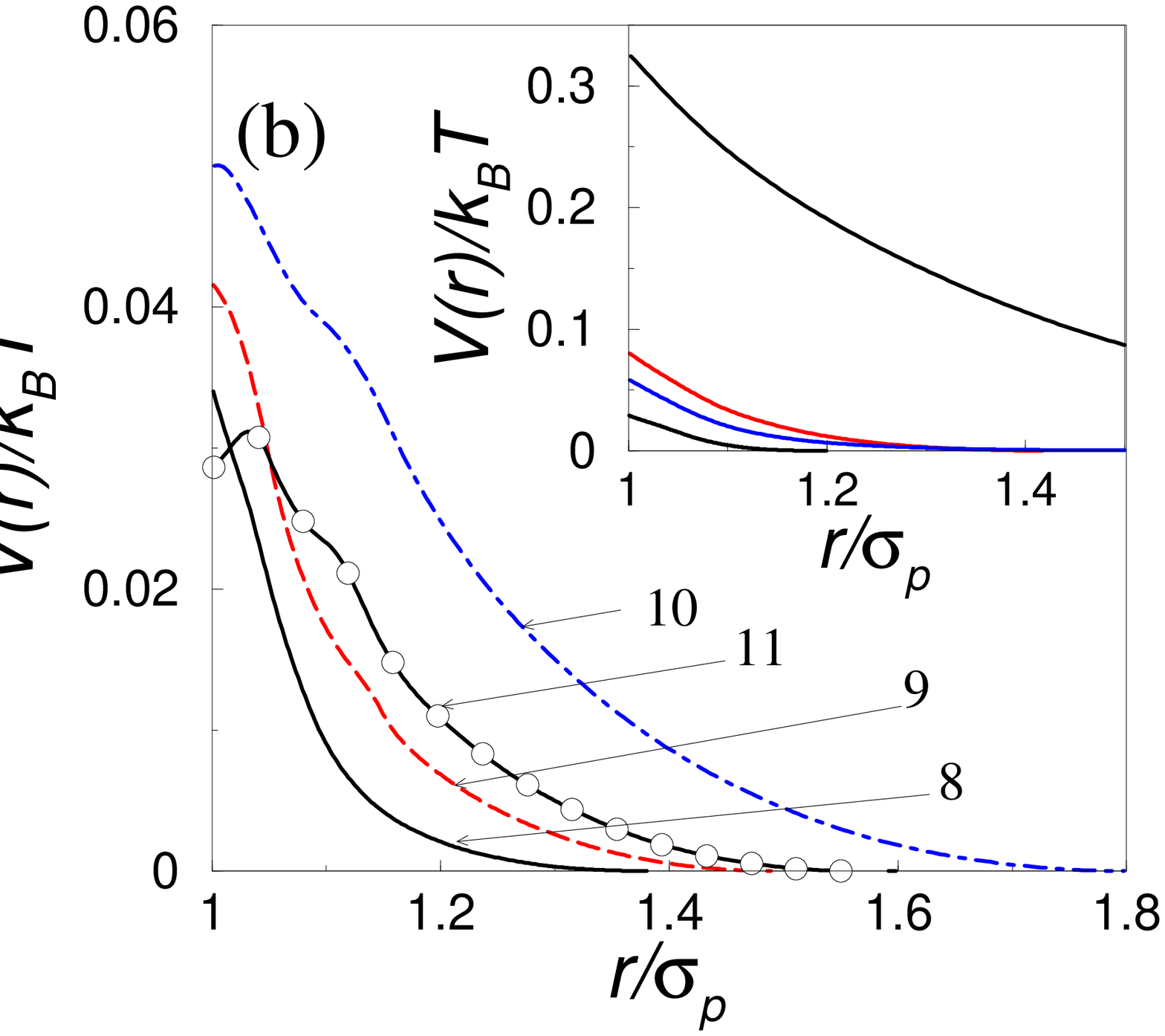} 
   \caption{The same as in Figure~\ref{dcmforcepotentialz10} but now 
   for protein charge (a) $Z=6$ and (b)$Z=15$. The run numbers are 
   placed next to corresponding curves. The result for run 1 is 4 
   times reduced in $y$-value to fit the $y$-axis scale. The inset in 
   (b) shows low salt concentrations, from top to bottom, runs 1, 4, 
   5, 7.}  \label{apotentialz6} 
\end{figure}

It is clear that the effective forces and potentials between two 
proteins will no longer be spherically symmetric within the DCM model. 
For example, three distinguishable mutual orientations of the two 
proteins are schematically outlined in Figure~\ref{dcmomega}a, 
corresponding to particular configurations of the Euler angles 
$\vec{\Omega}_1,\vec{\Omega}_2$ of the two proteins.  Nevertheless, 
our simulation results, presented in Figure~\ref{dcmomega}b, for the 
three orientations, show that the actual anisotropy of force is weak. 
However, this is no longer true for the Yukawa segment model 
\cite{AllahyarovDNA,stigter,delrow}, as shown in the  
 inset of Figure~\ref{dcmomega}b.  Within this model, the total 
effective interaction potential between a pair of protein spheres is 
given by 
\begin{equation}  
 U^{(YS)}(r) = \frac{1}{Z^2} \sum_{k,n=1}^{Z} 
   U^{(DLVO)} ( \mid {\vec r}_k - {\vec r}_n \mid) , 
\label{FYS} 
\end{equation}  
where 
${\vec r}_k$ and ${\vec r}_n$ represent the positions of the  point
unit charges of different proteins.  We emphasize that 
the aelotopic (or nonisotropic) interactions incorporated in our DCM 
differ from those considered, for example, in Ref.\cite{Benedek}, 
where $B_2$ is calculated for a set of hydrophobic {\it attractive} patches 
on protein surface.  Within our version of the DCM the third 
configuration in Fig.~\ref{dcmomega} (solid line), has the highest 
statistical weight of the three cases pictured (see 
Eqs.(\ref{potential},\ref{A})).  If, on the other hand, the point 
charges on the protein are replaced by attractive 
patches\cite{Benedek}, then the configuration with two points nearly 
touching (dot-dashed line in Figure~\ref{dcmomega}), is the 
statistically most favorable conformation. Similar arguments hold 
within a molecular model for site-specific short-range attractive 
protein-protein interactions,\cite{Vega,Anderson}. 
Results for distance-resolved forces within the DCM model are shown in 
Figure~\ref{dcmforcepotentialz10}a, for $Z=10$. When the salt 
concentration is less than $C_s\lesssim 0.2$ Mol/l, the results are 
similar to those of the SCM model: i.e. for low ionic strength, the 
force is repulsive, while for high ionic strength there is an 
attraction near contact followed by a repulsive barrier.  The 
distinguishing property of the DCM is the {\em nonmonotonicity} of the 
force with the increase of ionic strength. This, in turn, gives rise 
to the nonmonotonic behavior of the spherically averaged interaction 
potential $V(r)$ shown in Figure~\ref{dcmforcepotentialz10}b. 
This feature of $V(r)$ manifests itself in the following way in 
Figure~\ref{dcmforcepotentialz10}b: the potential is first strongly 
reduced as $C_s$ is increased, then its amplitude and range increase 
very significantly at intermediate concentrations ($C_s 
\simeq 1$ Mol/l), before it nearly vanishes at the highest salt 
concentrations. Note that $V(r)$ even becomes slightly attractive at 
contact ($r=\sigma_p$) for $C_s \simeq 2$ Mol/l. Similar effects are 
also observed for $Z=6$ and $Z=15$ (see Figure~\ref{apotentialz6}), 
suggesting that the effect is generic for discrete charge 
distributions. 
 
Once the effective potential $V(r)$ is known, it is straightforward to 
calculate the second osmotic virial coefficient using 
Eq.(\ref{eqB2}). In doing so, however, one should keep in mind that it 
is the total interaction that enters $B_2$. Real proteins also exhibit 
an additional short-range interaction, as seen, for example,  in 
experimental studies of the osmotic pressure and structural data for 
lysozyme\cite{Degiorgio}, or in fits to its phase-behavior\cite{Rose95}. 
This attraction stems from hydration forces, van-der-Waals 
interactions, and other molecular interactions that are, to a first 
approximation, independent of salt concentration. Hence, we have taken 
the expected short-range attraction between proteins into account by 
adding to the effective Coulomb potential in Eq.(\ref{eq2}), an 
additional ``sticky'' sphere potential of the Baxter 
form\cite{baxter}, 
\begin{equation} 
\frac {V_{SHS}(r)} {k_BT} = \cases  
{\infty & $ r \leq \sigma_p  $\cr 
   \ln \left [ {\frac{12 \tau \delta} {\sigma_p + \delta}}  \right ] &  
$ \sigma_p < r < \sigma_p + \delta$ ,\cr   
   0 & $r \geq \sigma_p + \delta$ \cr} 
\nonumber 
\end{equation}  
 with potential parameters $\delta = 0.02 
\sigma_p$ and $\tau = 0.12$, which  yield reasonable 
osmotic data for lysozyme solutions\cite{Rose95,Rosenbaum,Degiorgio} 
in the high salt concentration regime. This square well potential is 
isotropic by nature and ignores the directionality in hydrophobic 
attraction between proteins \cite{Haas,Sear}.  Short range attractions 
lead to ``energetic fluid'' behavior\cite{Loui01a}, where the 
crystallization is driven primarily by the details of the 
interactions, instead of being dominated by the usual entropic 
hard-core exclusions. This suggests that the directionality may be 
very important to details of the protein crystallization 
behavior\cite{Sear}.  However, for the physically simpler behavior 
of the virial coefficient, the directionality can be ignored as a 
first approximation.  For simplicity, we assume the parameter $\tau$ 
to be independent of electrolyte conditions, although a weak 
dependence based on experimental observations is reported in 
\cite{Rosenbaum,Anderson}. The addition of $V_{SHS}(r)$ strongly 
magnifies the nonmonotonicity of $B_2$ stemming from the nonmonotonic 
behavior of $V(r)$ near contact. 
 
\begin{figure} 
   \epsfxsize=13cm 
   \epsfysize=13cm 
~\hfill\epsfbox{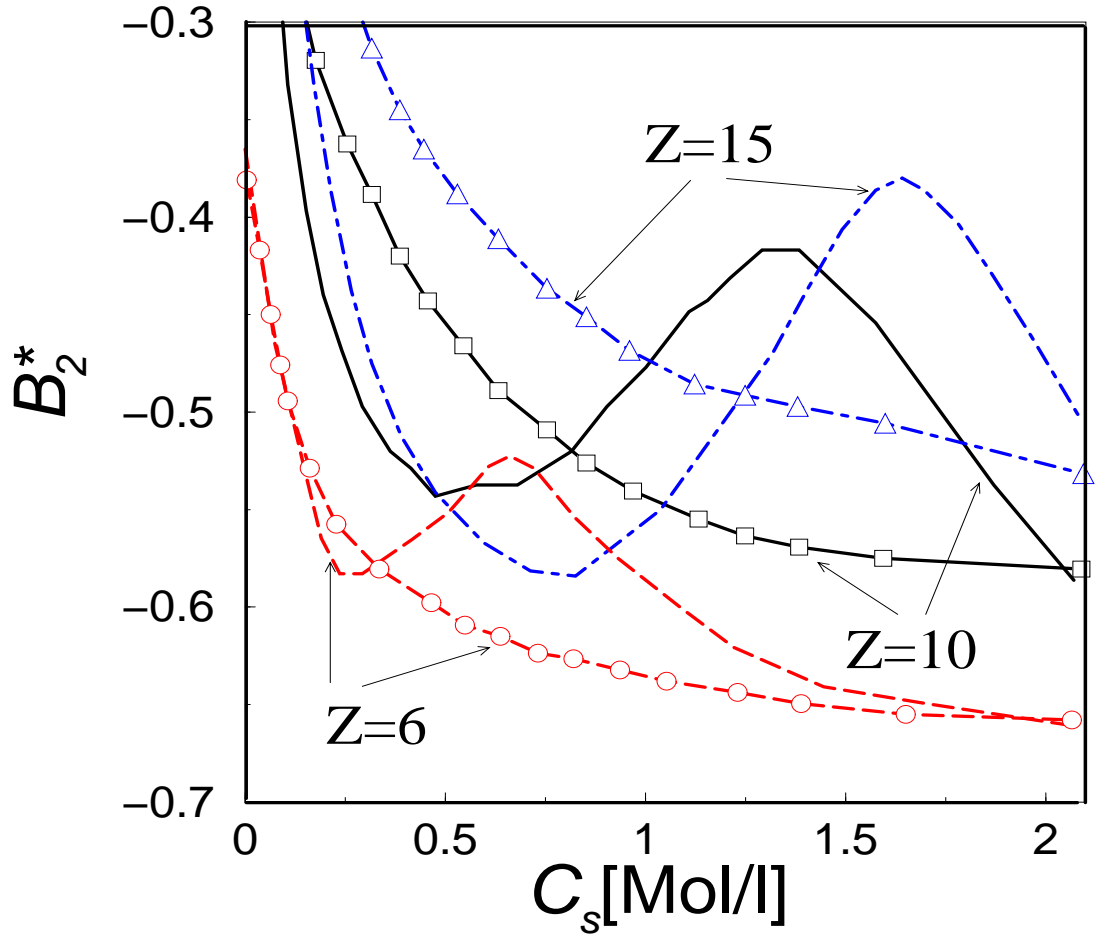}\hfill~ 
\caption{\label{B2} 
Normalized second virial coefficient $B_2^* = B_2/B_2^{HS}$ of a 
protein solution versus salt concentration $C_s$. Results are shown 
for protein charges $Z=6$ (dashed lines), $Z=10$ (solid lines) and 
$Z=15$ (dot-dashed lines). The lines with (without) symbols correspond 
to the SCM (DCM) model. Whereas the SCM virial coefficients decrease 
monotonically with increasing salt concentration, as expected from 
simple screening arguments, the DCM shows a marked {\em nonmonotonic} 
increase of $B_2$ at intermediate salt concentrations.} 
\end{figure}

 Results for $B_2^*$ as a function of salt concentration are shown in 
Figure~\ref{B2} for three different protein charges.  There is a 
considerable {\em qualitative} difference between the predictions of 
the SCM and the DCM models for the variation of $B_2^*$ with 
monovalent salt concentration $C_s$ for each protein charge $Z$. 
Whereas the SCM (dashed curves in Figure~\ref{B2}) predicts a 
monotonic decay of $B_2^*$ with $C_s$, the DCM leads to a markedly 
non-monotonic variation, involving an initial decay toward a minimum 
(salting-out) followed by a subsequent increase to a maximum 
(salting-in) and a final decrease at high $C_s$ values 
(salting-out). The location of the local minima shifts to higher/lower 
values of $C_s$ for larger/smaller protein charges $Z$. Thus for 
larger protein charge one needs a higher salt concentration to 
achieve the ``salting out'' conditions conducive to protein 
crystallization~\cite{protein3}.  Even though the effective Coulomb 
potential between proteins is small, with an amplitude only a few 
percent of the thermal energy $k_B T$, its effect on $B_2$ is 
dramatically enhanced by the presence of the strong short-range 
attractive Baxter potential.

We have also compared the effective potentials shown in figure 
\ref{B2} with a recently proposed scaling collapse of  
protein osmotic virial coefficients~\cite{protein3}.  This scaling 
effect, observed for a number of experimental 
conditions\cite{protein3}, can be explained with simple arguments 
based on Donnan equilibrium\cite{Warren}.  To lowest order, the 
effects of salt concentration and protein charge on $B_2$ are to 
subsumed in the following approximate scaling relation: 
\begin{equation}\label{Donnan} 
B_2^{(0)}=B_2 - Z^2/4C_s, 
\end{equation} 
where $B_2^{(0)}$ is the bare virial coefficient, independent of 
charge effects.  As shown, for example in Figure 1 of 
reference~\cite{Warren}, this simple relation hold remarkably well 
above a salt concentration of $C_s \approx 0.25 M$ for a wide range of 
experimental measurements of $B_2$ for lysozyme, which all tend to a 
plateau value of $B_2^{0}/B_2^{(HS)} \approx (-2.7 \pm 0.2)$.  One 
implication of this observed scaling is that the attractive 
interactions that govern $B_2^{(0)}$ are indeed roughly independent of 
salt concentrations above $C_s \approx 0.25 M$.  When we applied the 
same scaling procedure to our $B_2$ curves, a similar plateau develops 
for both the DCM and the SCM models, albeit with $B_2^{(0)}$ slightly 
less negative than that found in the experiments.  One could, of 
course, very easily match to experiments by adjusting the value of 
$\tau$, but to keep contact with our earlier work\cite{ourEPL}, we 
don't do so.  Clearly the scaling does bring the DCM and SCM $B_2$'s 
close together for a given $Z$, but for different $Z$ (related to 
solution pH), the scaling collapse is not as good as that seen in 
experiments, since we observe a larger $C_s$ before it sets in. 
Nevertheless, considering the high density of co and counter-ions in 
the simulation, it is remarkable that a simple Donnan theory based on 
ideal gas terms performs so well.

 The origin of the non-monotonic variation of $B_2^*$ with $C_s$ can 
be traced back to the subtle correlation effects which cause an 
enhancement of the effective Coulomb repulsion at intermediate salt 
concentrations in the DCM\@. These effects cannot be rationalized in 
terms of simple mean-field screening arguments \cite{Linse2}. The protein-microion 
correlations are of a different nature to those in the SCM, where they 
lead to a much more conventional, monotonic decay of $B_2$ with $C_s$, 
similar to that expected from a simple screening picture.

In order to gain further insight into the physical mechanism 
responsible for the unusual variation of the effective Coulomb 
potential and of $B_2$ with salt concentration in the DCM, we consider 
the influence of a second near-by protein on the microion distribution 
near protein-ion contact. We have computed the difference between 
``inner'' and ``outer'' shell microion contact densities for $Z=10$, 
as schematically illustrated in the inset to 
Figure~\ref{imbalance}. The local microion density is no longer 
spherically symmetric, due to the interference of the electric 
double-layers associated with the two proteins. The difference $\Delta 
\rho = \rho_{ in} - \rho_{ out}$ between the mean number of microions 
within a fraction of a spherical shell of radius $R=0.6 \sigma_p$ 
subtended by opposite $60^{\circ}$ cones, is plotted in 
Figure~\ref{imbalance} versus salt concentration. $\Delta \rho$ is 
always positive, indicating that microions ( mainly counterions) tend 
to cluster in the region between the proteins, rather than on the 
opposite sides.  This follows because they can lower the total 
electrostatic energy by being shared between two proteins.  However, 
there is a very significant difference in the variation of $\Delta 
\rho$ with salt concentration $C_s$, between the SCM and the DCM 
models. Both exhibit similar behavior for lower salt concentrations 
$C_s \leq 0.5$ Mol/l; for example, both show a small maximum around 
$0.2$ Mol/l. But for salt concentrations above $0.5$ Mol/l, the SCM 
predicts a monotonic decrease of $\Delta 
\rho$, while the DCM leads to a sharp peak in $\Delta \rho$ for $C_s 
\simeq 1$ Mol/l. This highly non-monotonic behavior clearly correlates 
with the non-monotonicity observed in 
Figs.~\ref{dcmforcepotentialz10},~\ref{apotentialz6} and~\ref{B2}. The 
basic mechanism can be summarized as follows: For the DCM, the excess 
number of microions between the two proteins leads to an excess 
entropic pressure or force, as demonstrated in Figure~\ref{fix12}, 
which is the origin of the increased {\em repulsion}\/ between 
proteins around $C_s = 1$ Mol/l.  The enhanced microion density arises 
from subtle crowded charge correlation effects that cannot easily be 
understood at a mean-field level.

\section{conclusion} 
 
In conclusion, we have calculated the effective interactions and the 
second osmotic virial coefficient $B_2$ of protein solutions 
incorporating the electrostatics within the ``primitive'' model of 
electrolytes.  In this way we include nonlinear screening, 
overscreening, and correlation effects missed within the standard 
Poisson-Boltzmann description. For discrete charge distributions, the 
interactions and related $B_2$ vary in a non-monotonic fashion for 
increasing ion strength while for the smeared charge model, a standard 
workhorse of colloidal physics, this effect is absent.  These
correlation induced effects 
are missed within non-linear PB theory, and similar coarse-graining 
techniques taken from the theory of colloids.  In addition to this, 
our simulations indicate {\it the necessity of taking entropic forces into 
account} when treating systems on the nanoscale.  These forces are 
believed to be essential in the salting-out effect \cite{protein2} and 
could lead to an attraction even between neutral globular 
proteins \cite{Wu}.

Our MD calculations can easily be extended to the more complex (pH 
dependent) charge patterns of realistic proteins\cite{Boye99}.  In 
fact, in some cases it may be easier to do a full MD simulation than 
to solve the non-linear PB equations in a very complicated geometry. 
We expect similar mechanisms to those found for the DCM to be active 
there, leading, for example, to an enhanced protein-protein repulsion 
at intermediate salt concentration.  Since the second osmotic virial 
coefficient determines much of the excess (non-ideal) part of the 
chemical potential of semi-dilute protein solutions, we expect the 
non-monotonicity of $B_2$ to have a significant influence on protein 
crystallization from such solutions in the course of a ``salting-out'' 
process. The non-monotonic behavior also suggests the possibility of 
an inverse, ``salting-in'' effect, whereby a reduction of salt 
concentration may bring $B_2$ into the ``crystallization 
slot''\cite{Geor94,Rose96}. The sensitivity of $B_2$ to 
ion-correlation effects may help explain the salt specificity of the 
Hofmeister series~\cite{protein2}. Finally, we stress that our 
non-monotonicity is qualitatively different from that observed for 
added non-adsorbing~\cite{Kulkarni,Eisenriegler} and 
absorbing~\cite{Zerenkova} polymers. 
 
\acknowledgments 
The authors are grateful to R. Piazza, I.L. Alberts, 
P.G. Bolhuis, G. Bricogne, J. Clarke, S. Egelhaaf, J.F. Joanny,
D. Rowan, R.Blaak and W.C.K. Poon for useful discussions, and to Schlumberger Cambridge 
Research and the Isaac Newton Trust for financial support.

\begin{table} 
\caption{Parameters used for the different simulation runs. $N_s$ is 
    the number of salt ion pairs in simulation box, $C_s$ is the 
salt concentration in Mol/l, the Debye screening length 
$r_D$ is defined by  Eqn.(\ref{debye}) 
 and $r_s=\left({\frac{3 V^{'}}{4 \pi (2N_s+2Z)}}\right)^{\frac{1}{3}}$ is the average  
 distance between salt ions for a given salt concentration.} 
\begin{tabular}{lcccc} 
Run &  $N_s$ & $C_s$ (Mol/l) & $r_D/\sigma_p$ & $r_s/ \sigma_p $ \\ 
\tableline 
1& $      0$         & $0$      & $0    $   & $0$       \\  
2& $ 125$    & $0.05$   & $0.34 $   & $0.39$    \\  
3& $ 250$    & $0.103$  & $0.24 $   & $0.31$     \\  
4& $ 500$    & $0.206$  & $0.17 $   & $0.25 $    \\  
5& $ 1000$   & $0.412$  & $0.12 $   & $0.2 $   \\  
6& $ 1500$   & $0.62$   & $0.1  $   & $0.17$    \\  
7& $ 2000$   & $0.824$  & $0.085$   & $0.16 $   \\  
8& $ 2500$   & $1.03$   & $0.077$   & $0.15 $   \\  
9& $ 3000$   & $1.24$   & $0.07 $   & $0.14 $   \\  
10& $ 4000$  & $1.65$   & $0.06 $   & $0.124 $    \\  
11& $ 5000$  & $2.061$  & $0.054$   & $0.118 $   \\  
\end{tabular}  
\end{table}


\begin{figure} 
   \epsfxsize=13cm 
   \epsfysize=13cm 
~\hfill\epsfbox{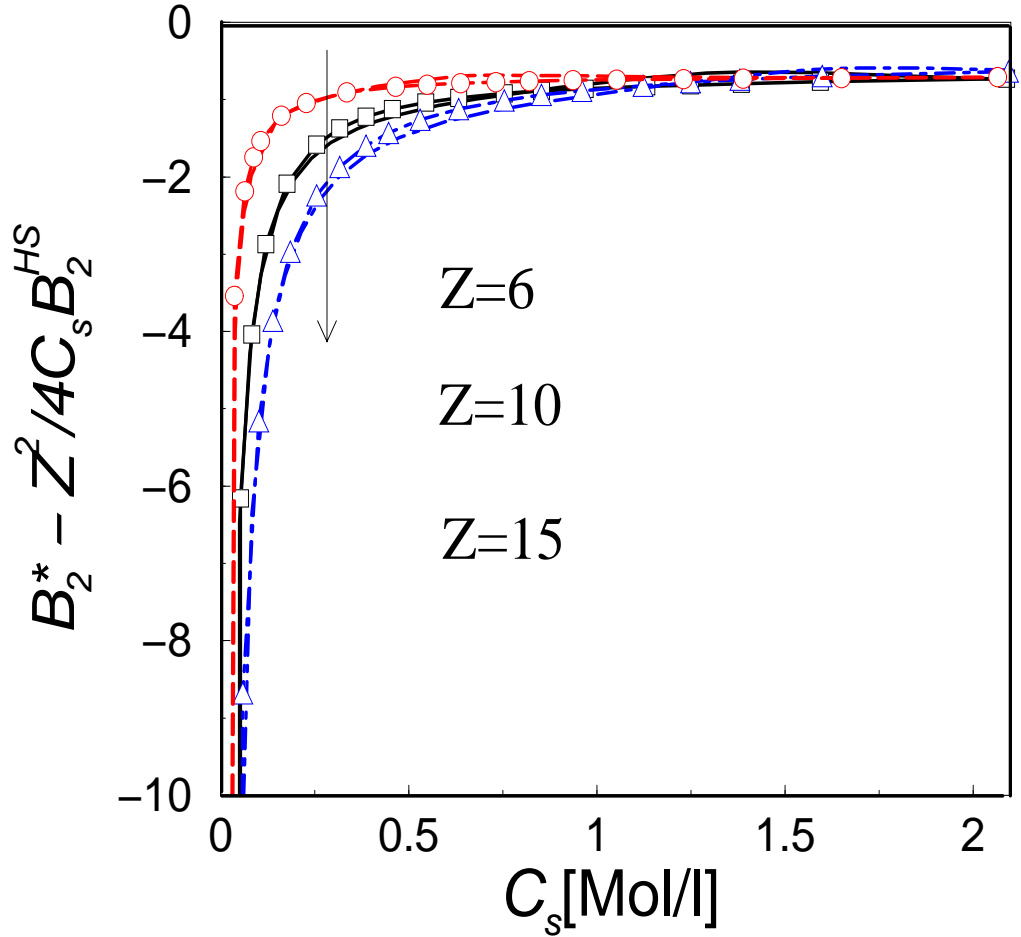}\hfill~ 
\caption{\label{B2new} 
The same runs as in Figure \ref{B2}, but now for the bare virial 
coefficient determined as $\frac {B_2 - Z^2/4C_s} {B_2^{(HS)}}$. The 
arrow is a guide for eye for the direction of increasing protein 
charge $Z$.  The scaling collapse at high $C_s$ has been related to a 
Donnan equilibrium effect\protect\cite{Warren}. Note that, on the
scale shown, the nonmonotonicity is hardly visible.} 
\end{figure} 
\begin{figure} 
   \epsfxsize=13cm 
   \epsfysize=13cm 
~\hfill\epsfbox{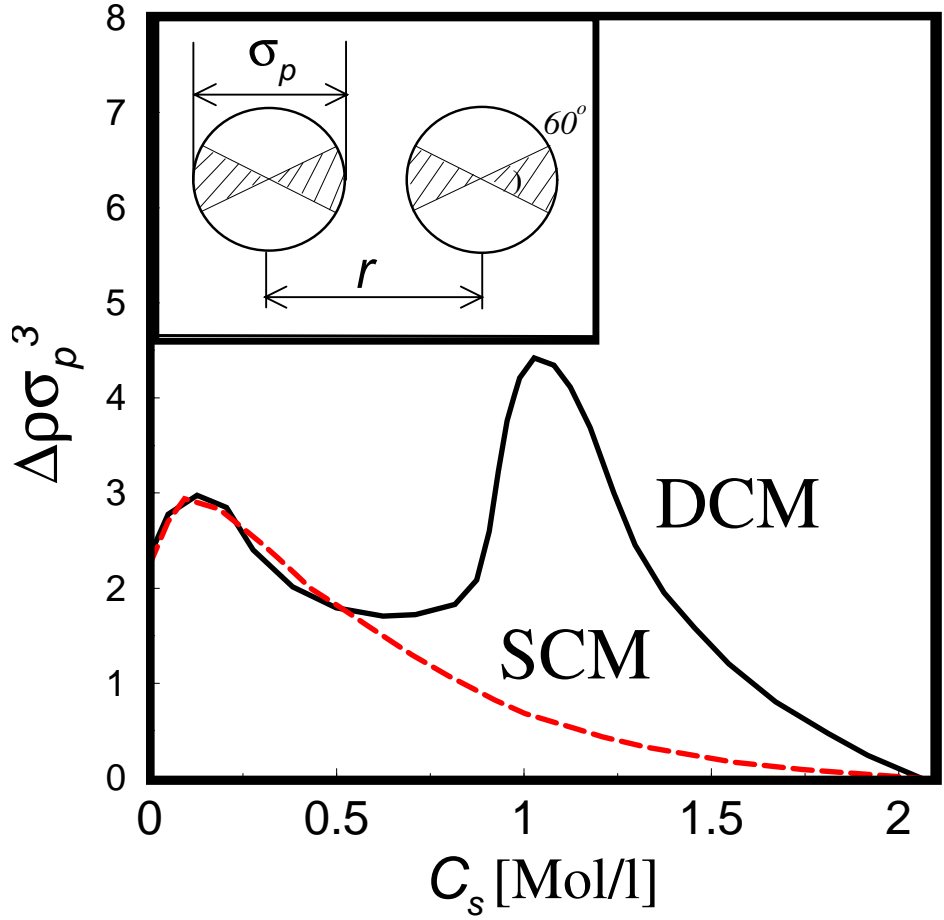}\hfill~ 
\caption{Difference in the microion density near contact $\Delta \rho$ versus 
salt concentration for protein charge $Z=10$ at a protein-protein 
separation of $r=1.2 \sigma_p$. The solid and dashed lines correspond 
to the DCM and SCM models respectively. The inset shows the angular 
range over which $\Delta \rho$ is averaged (see text).  The 
non-monotonic density profile for the DCM lies at the origin of the 
non-monotonic behavior seen for the forces, potentials, and virial 
coefficients calculated for this model.} 
\label{imbalance} 
\end{figure} 
\begin{figure} 
   \epsfxsize=15cm 
   ~\hfill\epsfbox{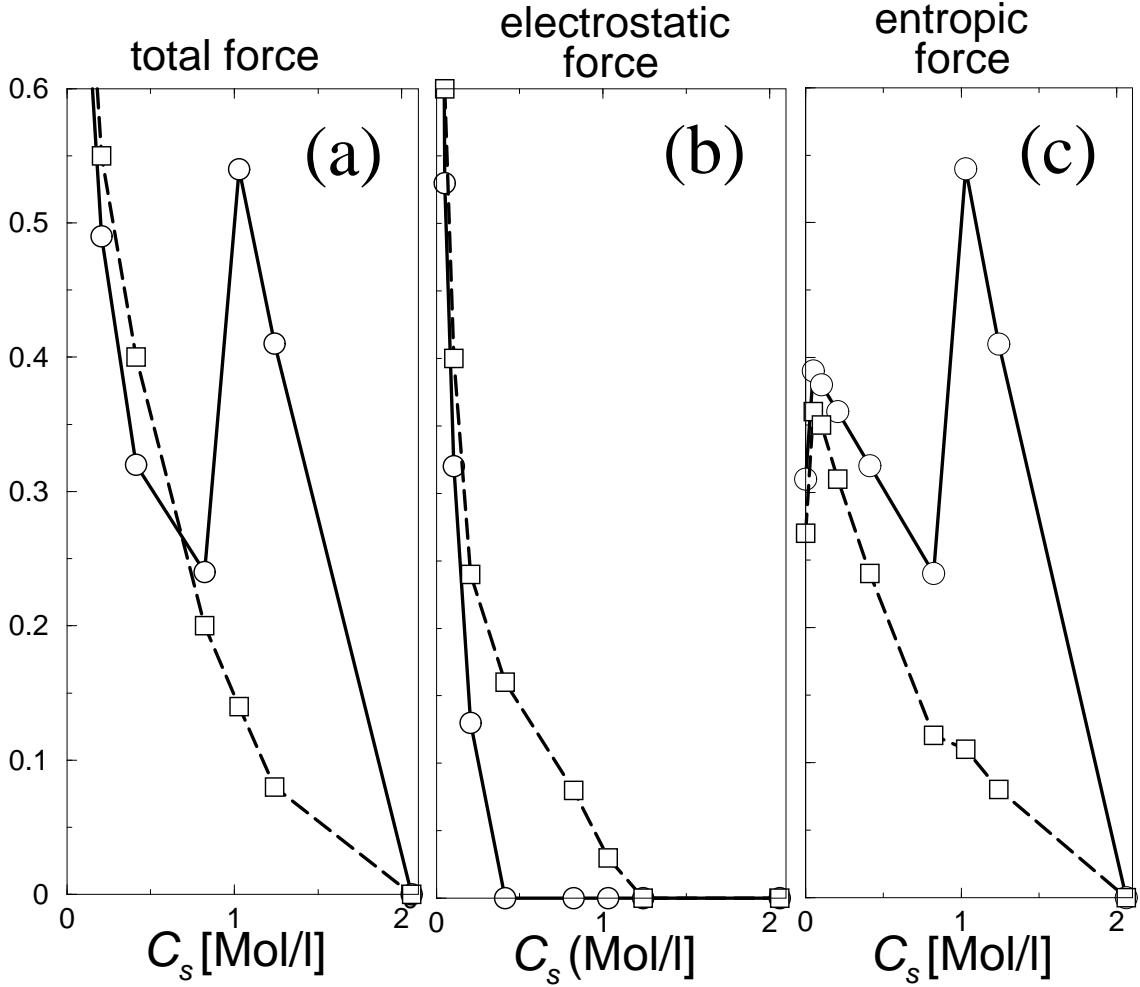}\hfill~ 
\caption{The protein-protein interaction force and its components at a 
   protein-protein separation $r= 1.2 \sigma_p$, in units of $k_BT/ 
   \lambda_B$, versus salt concentration $C_s$ for a charge of $Z=10$. 
   From left to right: $(a)$- total interaction force, $(b)$- 
   electrostatic component of interaction force, $(c)$- entropic 
   component of interaction force. Solid line- DCM and dashed line- 
   SCM results. This figure demonstrates that the difference between 
   the two models arises primarily from the contributions of the 
   entropic force.}  \label{fix12} 
\end{figure} 

\appendix  
\section{Multipole expansion of counterion density around protein in DCM} 
 
We expand the numerically calculated counterion density around a 
protein with charge $Z$ in a Laplace series of spherical harmonics: 
\begin{equation} 
\rho(r,\theta,\varphi) = \sum_{n,m}  { C_{nm}(r) P_n^m(\cos 
    (\theta) ) \cos(m\varphi) + S_{nm}(r) P_n^m(\cos(\theta) 
    )\sin(m\varphi) } 
\end{equation} 
The multipole spherical expansion coefficients (MSEC) are calculated
during an MD simulation via 
\begin{eqnarray} 
C_{00}& =& \langle \sum_i \delta(\vec r - \vec r_i) \rangle \nonumber ,\\ 
C_{10}& =& \langle \sum_i \delta(\vec r - \vec r_i) \cos(\theta_i) 
\rangle \nonumber ,\\ 
C_{11}& =& \langle \sum_i \delta(\vec r - \vec r_i) \sin(\theta_i) 
\cos(\varphi_i) \rangle \nonumber,\\ 
S_{11}& =& \langle \sum_i \delta(\vec r - \vec r_i) \sin(\theta_i) 
\sin(\varphi_i) \rangle \nonumber,\\ 
C_{20}& =& \langle \sum_i \delta(\vec r - \vec r_i) \frac{3 
    \cos^2(\theta_i)-1} {2} \rangle \nonumber,\\ 
C_{21}& =& \langle \sum_i \delta(\vec r - \vec r_i) 3 
  \sin(\theta_i)\cos(\theta_i) \cos(\varphi_i) \rangle \nonumber,\\ 
S_{21}& =& \langle \sum_i \delta(\vec r - \vec r_i) 3 
  \sin(\theta_i)\cos(\theta_i) \sin(\varphi_i) \rangle \nonumber,\\ 
C_{22}& =& \langle \sum_i \delta(\vec r - \vec r_i) 3 
\sin^2(\theta_i)\cos(2\varphi_i) \rangle \nonumber,\\ 
S_{22}& =& \langle \sum_i \delta(\vec r - \vec r_i) 3 
\sin^2(\theta_i)\sin(2\varphi_i) \rangle \nonumber, 
\end{eqnarray} 
where $i$ runs over counterions. 
 
\begin{figure} 
   \epsfxsize=10cm 
   \epsfysize=11cm 
\epsfbox{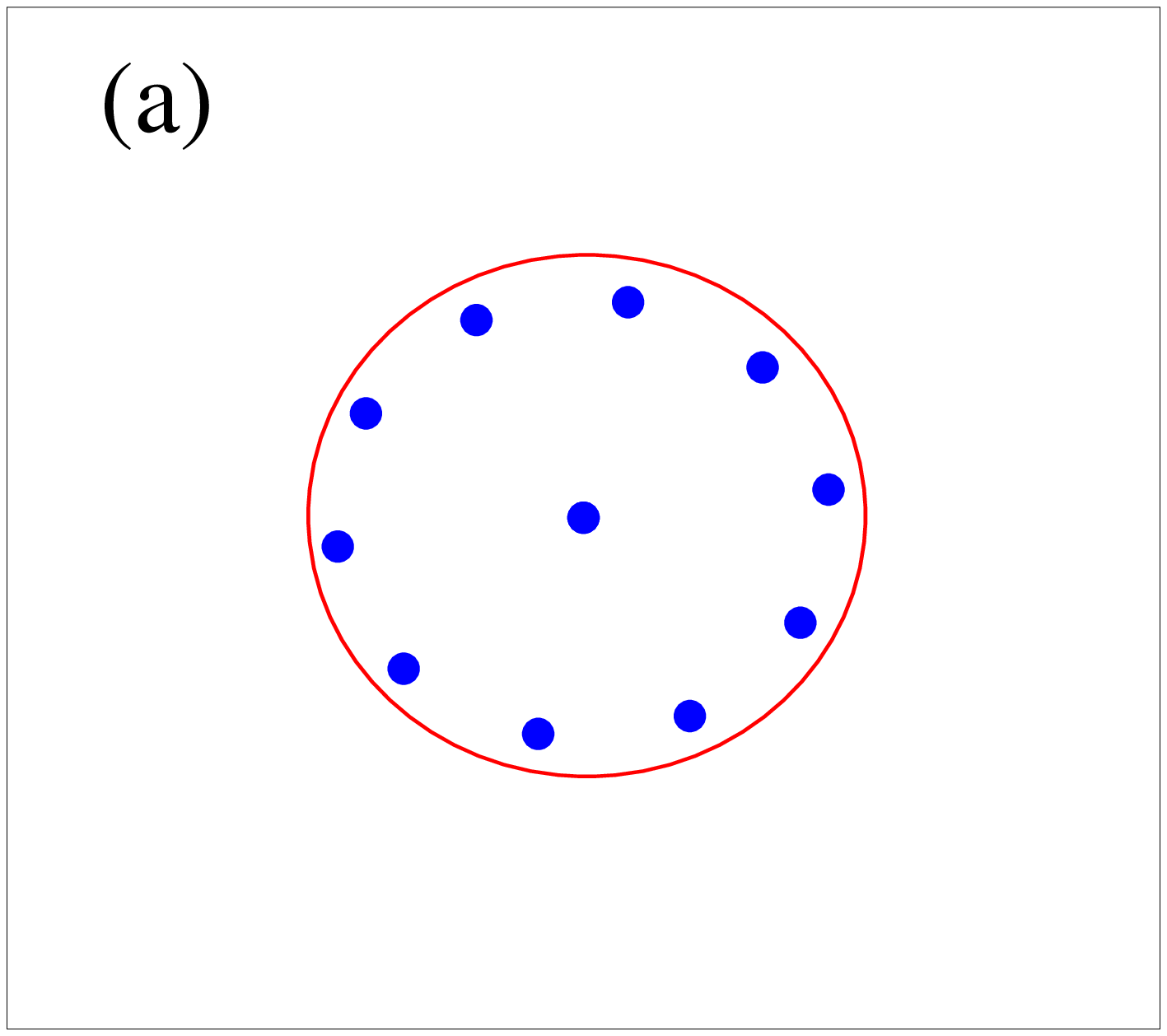} \vskip -2cm
   \epsfxsize=10cm 
   \epsfysize=11cm 
\epsfbox{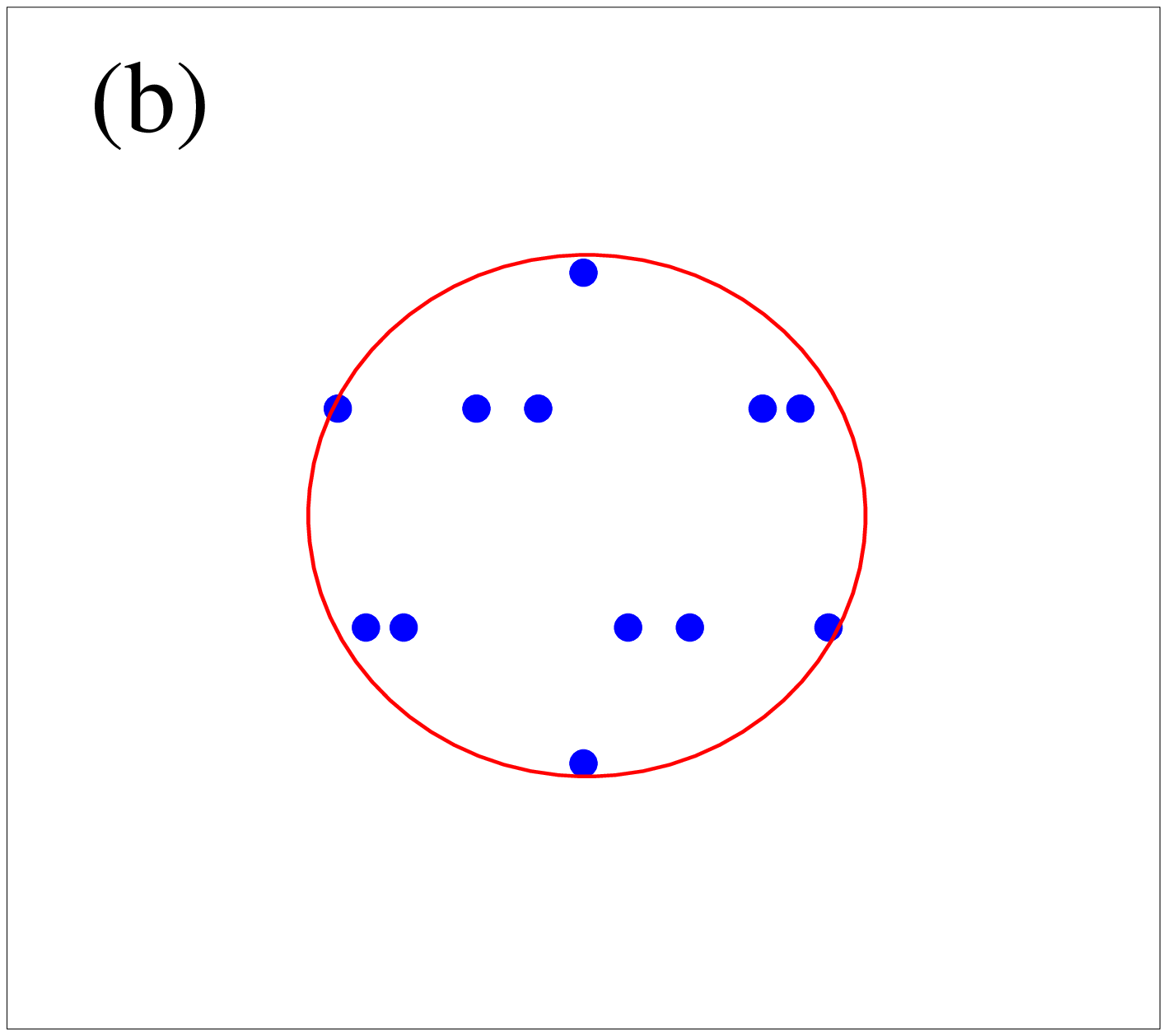}\hfill~ 
\vspace{1cm}
\caption{A schematic diagram for the $xy$ plane (a) and $xz$ plane (b)
  projections of the protein surface
  charge distribution in DCM model for protein charge $Z=12$.} 
     \label{A_multipole_fig1} 
\end{figure}

The protein charge is chosen to be $Z=12$, so that the rotation 
symmetry axis through two surface charges has a fivefold 
symmetry.  The $xz$ and $xy$ plane projections of the protein charge 
pattern are shown in Figure \ref{A_multipole_fig1}.

\begin{figure} 
  \epsfxsize=12cm 
   \epsfysize=12cm 
~\hfill\epsfbox{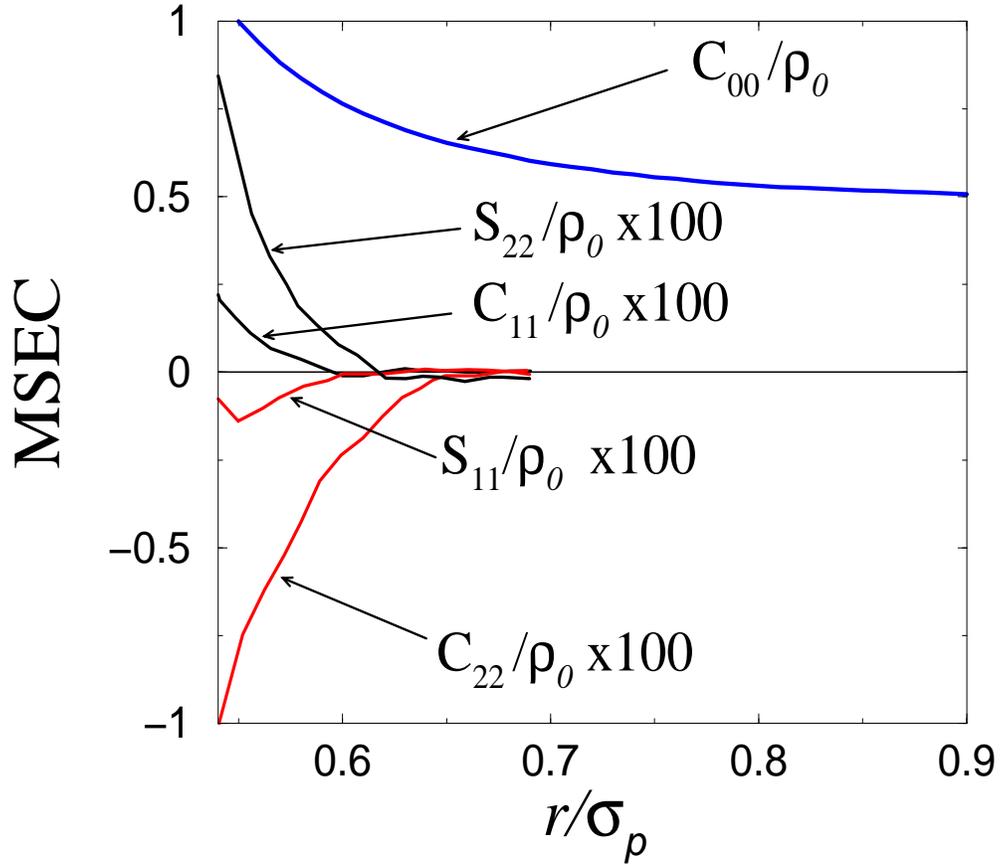}\hfill~ 
\caption{Five nonzero multipole spherical expansion 
  coefficients (MSEC) of counterion density $\rho_{+}(r)$ for the DCM
  with a protein charge $Z=12$ and
  salt concentration $C_s=0.05$ Mol/l. Note that the higher 
  order expansion coefficients are magnified (x100).} 
     \label{A_7} 
\end{figure}

The variation of the nonzero MSEC versus distance 
from the protein center  are shown in Figure \ref{A_7}. 
 
\end{document}